\documentclass{siamltex}
\usepackage{euscript,amsmath,amssymb,amsfonts,graphicx,bm}
  \usepackage{epstopdf}
  \usepackage{graphics} 
 \usepackage[colorlinks=true]{hyperref}
 \hypersetup{urlcolor=blue, citecolor=red}
 \usepackage[latin1]{inputenc}
\newcommand{\calU}{{\mathcal U}}
\newcommand{\calG}{{\mathcal G}}
\newcommand{\calJ}{{\mathcal J}}
\newcommand{\calC}{{\mathcal C}}
\newcommand{\calF}{{\mathcal F}}
\newcommand{\R}{{\mathbb R}}

\newcommand{\X}{\mathbf{X}}
\renewcommand{\P}{\mathbb{P}}
\newcommand{\x}{\mathbf{x}}

\newcommand{\z}{\mathbf{z}}
\newcommand{\y}{\mathbf{y}}
\newcommand{\e}{{\mathrm e}}
\newcommand{\E}{{\mathbb E}}
\newcommand{\pcb}[1]{\textcolor{black}{#1}}
\newcommand{\n}{\mathbf n}

\renewcommand{\P}{\mathbb P}

\newcommand{\hxi}{{\bm \xi}}

\begin{document}

\title{The 3D narrow capture problem for traps with semipermeable interfaces}

\author{Paul C. Bressloff\thanks{Department of Mathematics, University of Utah, Salt Lake City, UT 84112
USA ({\tt bressloff@math.utah.edu})} }

\date{today}

 \maketitle

\begin{abstract} 
In this paper we analyze the narrow capture problem for a single Brownian particle diffusing in a three-dimensional (3D) bounded domain containing a set of small, spherical traps. The boundary surface of each trap is taken to be a semipermeable membrane. That is, the continuous flux across the interface is proportional to an associated jump discontinuity in the probability density. The constant of proportionality is identified with the permeability $\kappa$. In addition, we allow for discontinuities in the diffusivity and chemical potential across each interface; the latter introduces a directional bias. We also assume that the particle can be absorbed (captured) within the interior of each trap at some Poisson rate $\gamma$. In the small-trap limit, we use matched asymptotics and Green's function methods to calculate the splitting probabilities and unconditional MFPT to be absorbed by one of the traps. However, the details of the analysis depend on how various parameters scale with the characteristic trap radius $\epsilon$. Under the scalings $\gamma=O(1/\epsilon^2)$ and $\kappa=O(1/\epsilon)$, we show that the semipermeable membrane reduces the effective capacitance $\calC$ of each spherical trap compared to the standard example of totally absorbing traps. The latter case is recovered in the dual limits $\gamma \rightarrow \infty$ and $\kappa \rightarrow \infty$, with $\calC$ equal to the intrinsic capacitance of a sphere, namely, the radius. We also illustrate how the asymptotic expansions are modified when $\gamma=O(1/\epsilon)$ (slow absorption) or $\kappa =O(1)$ (low permeability). Finally, we consider the unidirectional limit in which each interface only allows particles to flow into a trap. The traps then act as partially absorbing surfaces with a constant reaction rate $\kappa$. Combining asymptotic analysis with the encounter-based formulation of partially reactive surfaces, we show how a generalized surface absorption mechanism (non-Markovian) can be analyzed in terms of the capacitances $\calC$. We thus establish that a wide range of narrow capture problems can be characterized in terms of the effective capacitances of the traps.

\end{abstract}

\section{Introduction}

An important problem in the mathematics of diffusion is analyzing transport through a semipermeable interface. Interfaces of this form occur in a wide range of processes, including  lipid bilayers regulating molecular transport in biological cells \cite{Philips12,Alberts15,Bressloff21,Nik21}, chemical and electrical gap junctions \cite{Evans02,Connors04,Good09,Bressloff16}, artificial membranes for reverse osmosis  \cite{Li10,Rubinstein21}, and animal migration in heterogeneous landscapes \cite{Beyer16,Assis19,Kenkre21}. Moreover, a variety of complex porous media are modeled in terms of multiple semipermeable interfaces and heterogeneous diffusivities \cite{Grebenkov10,Hahn12,Carr16,Aho16,Moutal19,Farago20,Alemany22}. The classical approach to incorporating a semipermeable interface into the diffusion equation is to impose flux continuity across the interface, with the flux proportional to an associated jump discontinuity in the concentration; the constant of proportionality represents the permeability. One method for deriving the semipermeable boundary conditions is to consider a thin membrane and to apply statistical thermodynamical principles \cite{Kedem58,Kedem62,Kargol96}. This leads to the so-called Kedem-Katchalsky equations, which also allow for discontinuities in the diffusivity and chemical potential across the interface. The latter introduces a directional bias.

In this paper we consider the problem of a single Brownian particle diffusing in a three-dimensional (3D) bounded domain $\Omega \subset \R^3$ containing a set of $N$ small, spherical traps $\calU_j$, $j=1,\ldots,N$. The boundary $\partial \calU_j$ of the $j$th trap is taken to be a semipermeable membrane with permeability $\kappa_j$ and a directional bias $\alpha_j\in [0,1]$. Diffusive flux is preferentially directed into the trap if $\alpha_j\in [0,1/2)$ and out of the trap if $\alpha_j\in (1/2,0]$; the unbiased case is $\alpha_j=1/2$.
We also allow for the diffusivity $D_j$ within a trap to differ from the bulk diffusivity $D$. Finally, we assume that the particle can be absorbed (captured) within the interior of each trap at a Poisson rate $\gamma$. In the small-trap limit, $|\calU_j| =\epsilon^3|\Omega|$ with $0<\epsilon \ll 1$, we use matched asymptotics and Green's function methods to solve the resulting narrow capture problem  \cite{Cheviakov11,Chevalier11,Coombs15,Bressloff15,Lindsay15,Lindsay17,Bressloff21B,Bressloff22}. In particular, we calculate the splitting probability $\pi_k(\x_0)$ to be captured by the $k$th trap, given that the particle started in the bulk domain at a position $\x_0$, and the corresponding unconditional MFPT $T(\x_0)$ for absorption. We investigate how these quantities depend on the model parameters, and consider various limiting cases. 

In the dual limits $\gamma \rightarrow \infty$ and $\kappa_j\rightarrow \infty$, $j=1,\ldots,N$, the interfaces $\partial \calU_j$ become completely permeable and the particle is immediately absorbed on entering a trap. Our model then reduces to the standard example of a 3D narrow capture problem, in which the surface of each trap acts as a totally absorbing surface $\partial \calU_j$. We show that in this limiting case our expressions for $\pi_k(\x_0)$ and $T(\x_0)$ are equivalent to previously obtained results \cite{Coombs15,Bressloff21B}. In particular, the leading order contribution to $\pi_k(\x_0)$ is an $O(1)$ constant, whereas the leading order contribution to $T(\x_0)$ is an $O(1/\epsilon)$ constant. On the other hand, for finite $\gamma$ and $\kappa_j$, the details of the asymptotic analysis depend on how these parameters scale with the size of the traps. Under the scalings $\gamma=O(1/\epsilon^2)$ and $\kappa=O(1/\epsilon)$, we show that the contributions from the semipermeable membranes are encapsulated by an effective reduction in the diffusive capacitance $\calC_j$ of each spherical trap compared to the standard example of totally absorbing traps, for which $\calC_j$ is the trap radius. We then explore how the asymptotic expansions are modified when $\gamma=O(1/\epsilon)$ (slow absorption) or $\kappa =O(1)$ (low permeability). In particular, we show that the $O(\epsilon)$ contribution to $\pi_k(\x_0)$ and the $O(1)$ contribution to $T(\x_0)$ are also now constants involving effective capacitances. 

Another important limiting case is obtained by taking $\alpha_j \rightarrow 0$ for finite $\kappa_j$, which means that once the particle enters a trap it can never escape (no outward flux). Each trap thus acts as a partially absorbing surface $\partial \calU_j$ with a Robin boundary condition. The corresponding reaction rate is determined by $\kappa_j$. We have recently analyzed a generalized version of the narrow capture problem with partially absorbing surfaces \cite{Bressloff22a} by combining matched asymptotic analysis with an encounter-based formulation of diffusion-mediated surface reactions \cite{Grebenkov20a,Grebenkov22,Bressloff22b,Bressloff22c}. The latter considers the joint probability density or generalized propagator for particle position and the so-called boundary local time, which characterizes the amount of time that a Brownian particle spends in the neighborhood of a point on a totally reflecting boundary. The effects of surface reactions are then incorporated via an appropriate stopping condition for the boundary local time. In particular, Robin boundary conditions are recovered in the special case of an exponential law for the stopping local times. In Ref. \cite{Bressloff22a} we derived an asymptotic expansion of the generalized propagator in the case of diffusion in an unbounded domain $\R^3$. (The traps were taken to be small compared to the minimum distance between any pair of traps.) One difference from a bounded domain $\Omega$ is that the splitting probabilities are $O(\epsilon)$, since 3D diffusion is transient rather than recurrent. Our main result was to show that for a general surface reaction, the leading order terms in the asymptotic expansion are characterized by an effective renormalization of the capacitance (radius) of the form $\calC_j \rightarrow \calC_j-\widetilde{\Psi}(1/\calC_j)$, where $\widetilde{\Psi}$ is the Laplace transform of the stopping local time distribution. In the final part of the paper we generalize this result to narrow capture in bounded domains, thus establishing that a wide range of narrow capture problems can be analzyed in terms of effective capacitances. (Alternative methods for analyzing the encounter-based narrow capture problem are developed in Ref. \cite{Grebenkov22a}.)

The structure of the paper is as follows. In section 2, we write down the forward reaction-diffusion equations for the narrow capture problem with semipermeable traps, define the splitting probabilities and MFPT, and discuss various limiting cases. We then derive the boundary value problems (BVPs) for these quantities by constructing the corresponding backward reaction-diffusion equations. The main nontrivial step is determining the adjoint semipermeable boundary conditions. In section 3 we carry out the matched asymptotic analysis of the splitting probabilities for $\gamma=O(1/\epsilon^2)$ and $\kappa=O(1/\epsilon)$, and derive an explicit expression for the effective capacitances $\calC_j$. In section 4 we develop the corresponding analysis of the unconditional MFPT. The explicit example of a triplet of traps in the unit sphere is presented in section 5, and the alternative forms of scaling are considered in section 6. Finally, in section 7, we consider the narrow capture problem with partially absorbing surfaces.

\section{The 3D narrow capture problem}
\begin{figure}[b!]
\centering
\includegraphics[width=12cm]{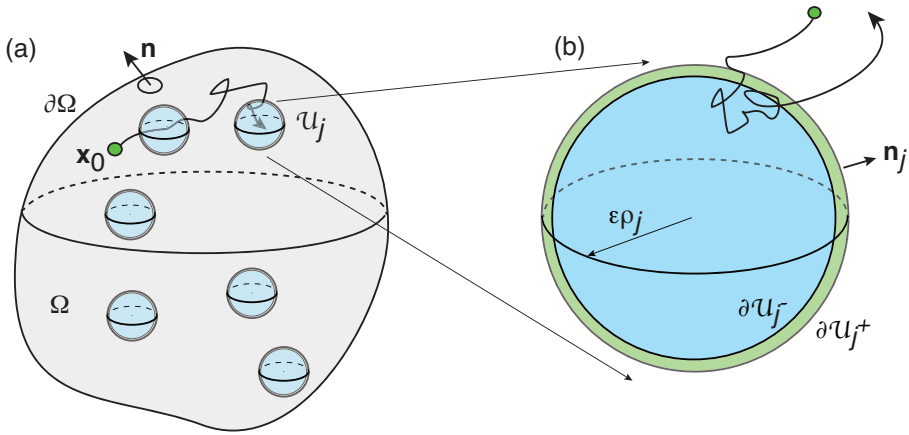} 
\caption{(a) A particle diffuses in a bounded domain $\Omega\subset \R^3 $ containing $N$ partially absorbing traps $\calU_j$, $j=1,\ldots,N$. Whenever the particle is within the trap domain $\calU_j$, it can be absorbed at a rate $\gamma$.The outward unit normals of $\partial \Omega$ and $\partial \calU_j$ are denoted by $\n$ and $\n_j$, respectively. (b)) The boundary $\partial \calU_j$ of the $j$th trap acts as a semipermeable interface. }
\label{fig1}
\end{figure}

Consider a set of partially absorbing traps $\calU_k\subset \Omega$, $k=1,\ldots,N$, in a bounded search domain $\Omega \subset \R^3$ and set $\bigcup_{k=1}^N \calU_k=\calU_a$, see Fig. \ref{fig1}. Whenever the particle is within $\calU_k$ it can be absorbed (react) at a rate $\gamma$. Each trap is taken to be much smaller than $\Omega$, that is, $|\calU_j|\sim \epsilon^3 |\Omega|$ with $\calU_j\rightarrow \x_j\in \Omega$ uniformly as $\epsilon \rightarrow 0$, $j=1,\ldots,N$. In addition, the traps are assumed to be well separated with $|\x_i-\x_j|=O(1)$, $j\neq i$, and $\mbox{dist}(x_j,\partial \Omega)=O(1)$ for all $i=1,\ldots,N$. For concreteness, we take each trap to be a 3-sphere of radius $\epsilon \rho_j$: $\calU_i=\{\x \in \Omega, \ |\x-\x_i|\leq \epsilon \rho_i\}$. Finally, the $j$th trap boundary $\partial \calU_j$  is treated as a semipermeable interface with $\partial \calU_j^+ $ ($\partial \calU_j^-$) denoting the side approached from outside (inside) $\calU_j$, see Fig. \ref{fig1}(b). The particle flux across each interface is continuous but there is a jump discontinuity in the concentration. Throughout the paper we will fix the length and time scales by taking $L\equiv |\Omega|^{1/3} =O(1)$ and $\tau\equiv L^2/D=O(1)$.

\subsection{Forward reaction-diffusion equations} Let $p(\x,t|\x_0)$ be the probability density that at time $t$ a particle is at $\X(t)=\x$, having started at position $\x_0$. We will set $p=q_j$ for all $\x\in \calU_j$ so that 
\begin{subequations} 
\label{master}
\begin{align}
	\frac{\partial p(\x,t|\x_0)}{\partial t} &= D\nabla^2 p(\x,t|\x_0), \ \x\in \Omega\backslash \calU_a,\\
	\frac{\partial q_j(\x,t|\x_0)}{\partial t} &= D_j\nabla^2 q_j(\x,t|\x_0) -\gamma q_j(\x,t|\x_0),\ \x\in \calU_j,
	\end{align}
together with the semipermeable boundary conditions
\begin{align}
 D\nabla p(\x^+,t|\x_0)\cdot \n_j &= D_j\nabla q_j(\x^-,t|\x_0)\cdot \n_j\equiv J_j(\x,t|\x_0) \\
	J_j(\x,t|\x_0)&=\kappa_j [(1-\alpha_j )p(\x^+,t|\x_0)- \alpha_jq_j(\x^-,t|\x_0)] ,\quad \x  \in \partial \calU_j ,
\end{align}
and the exterior boundary condition
\begin{equation}
\nabla p\cdot \n=0,\ \x \in \partial \Omega .
\end{equation}
\end{subequations}
Here $D$ and $D_j$ are the diffusivities in $\Omega\backslash \calU_a$ and $\calU_j$, respectively, $\n$ is the outward unit normal at a point on $\partial \Omega$, $\n_j$ is the outward unit normal at a point on $\partial \calU_j$, $J_j(\x,t)$ is the continuous inward flux across the point $\x\in \partial \calU_j$, $\kappa_j$ is the permeability of the interface $\partial \calU_j$, and $\alpha_j\in [0,1]$ specifies a directional bias with $\alpha_j=1/2$ the unbiased case. Equations (\ref{master}c,d) are one version of the well-known Kedem-Katchalsky (KK) equations \cite{Kedem58,Kedem62,Kargol96}. We will assume that the particle starts at some point $\x_0\in \Omega\backslash \calU_a$.

The absorption flux within the $k$-th trap at time $t$ is 
\begin{align}
\label{calJ}
	\calJ_k(\x_0,t)&= \gamma\int_{\calU_k} q_k(\x,t|\x_0)d\x,\ k = 1,\ldots,N.
	\end{align}
Hence, the splitting probability that the particle is eventually captured by the $k$-th trap is
\begin{equation}
\label{split}
\pi_k(\x_0)=\int_0^{\infty}\calJ_k(\x_0,t')dt' =\widetilde{\calJ}_k(\x_0,0),
\end{equation}
where $\widetilde{\calJ}_k(\x_0,s)$ denotes the Laplace transform of $\calJ_k(\x_0,t)$.
Let $Q(\x_0,t)$ denote the survival probability that the particle hasn't been absorbed by a trap in the time interval $[0,t]$, having started at $\x_0$:
\begin{align}
\label{Q1}
 Q(\x_0,t)=\int_{\Omega}p(\x,t|\x_0)d\x&=\int_{\Omega\backslash \calU_a}p(\x,t|\x_0)d\x+\sum_{k=1}^N\int_{\calU_k}q_k(\x,t|\x_0)d\x. \end{align}
Differentiating both sides of this equation with respect to $t$ and using equations (\ref{master}a,b) implies that
\begin{align}
 \frac{\partial Q(\x_0,t)}{\partial t}&=D\int_{\Omega\backslash \calU_a}\nabla^2p(\x,t|\x_0)d\x +\sum_{k=1}^N\int_{\calU_k}\left [D_k\nabla^2 q_k(\x,t|\x_0)-\gamma q_k(\x,t|\x_0)\right ]d\x\nonumber \\
 &=-D\sum_{k=1}^N \int_{\partial \calU_k}\nabla p\cdot \n_k d\sigma+\sum_{k=1}^N D_k \int_{\partial \calU_k}\nabla q_k\cdot \n_k d\sigma \nonumber \\
 &\quad  -\gamma \sum_{k=1}^N \int_{ \calU_k}q_k(\x,t|\x_0)d\x =-  \sum_{k=1}^N \calJ_k(\x_0,t),
\label{Q2}
\end{align}
where we have used the current conservation condition in equation (\ref{master}c).
Laplace transforming equation (\ref{Q2}) and imposing the initial condition $Q(\x_0,0)=1$ gives
\begin{equation}
\label{QL}
s\widetilde{Q}(\x_0,s)-1=- \sum_{k= 1}^N \widetilde{\calJ}_k(\x_0,s).
\end{equation}
In the case of a bounded domain $\Omega$, the particle is eventually absorbed by one of the traps with probability one, which means that
$\lim_{t\rightarrow \infty}Q(\x_0,t)=\lim_{s \rightarrow 0}s\widetilde{Q}(\x_0,s) =0$. Hence, $\sum_{k= 1}^N \widetilde{\calJ}_k(\x_0,s)=\sum_{k=1}^N\pi_k(\x_0)=1$. The normalized flux $\calJ_k(\x_0,t)/\pi_k(\x_0)$ is the conditional FPT density for absorption by the $k$th trap. Hence, the $k$th conditional MFPT is
\begin{equation}
T_k(\x_0)=\frac{1}{\pi_k(\x_0)}\int_0^{\infty}t\calJ_k(\x_0,t)dt=-\frac{1}{\pi_k(\x_0)}\left . \frac{\partial}{\partial s}\widetilde{\calJ}_k(\x_0,s)\right |_{s=0}.
\end{equation}
The corresponding unconditional MFPT is
\begin{equation}
\label{Tuncon}
T(\x_0)\equiv \sum_{k=1}^N \pi_k(\x_0)T_k(\x_0)=-\left . \frac{\partial}{\partial s}\sum_{k=1}^N \widetilde{\calJ}_k(\x_0,s)\right |_{s=0}=\widetilde{Q}(\x_0,0).
\end{equation}
Note that if the domain $\Omega$ were unbounded then $T(\x_0)$ would be infinite.

Finally, note that the Laplace transformed fluxes $\widetilde{\calJ}_j(\x_0,s)$ and $\widetilde{J}_j(\x,t|\x_0)$ can be related as follows. Integrating equation (\ref{master}b) with respect to $\x\in \calU_j$ and multiplying by $\gamma$ gives
\begin{align}
\frac{\partial \calJ_j(\x_0,t)}{\partial t}&= \gamma D_j\int_{\calU_j}\nabla^2 q_j(\x,t|\x_0) d\x-\gamma \calJ_j(\x_0,t)\nonumber \\
&= \gamma D_j\int_{\partial \calU_j}\nabla q_j(\x,t|\x_0) \cdot \n_j d\x-\gamma \calJ_j(\x_0,t)\nonumber \\
&=\gamma \int_{\partial \calU_j} J_j(\x,t|\x_0) d\x-\gamma \calJ_j(\x_0,t).
\end{align}
Laplace transforming this equation with respect to $t$ and using the initial condition $\calJ_j(\x_0,0)=0$ for $\x_0\notin \calU_j$, we have
\begin{equation}
(\gamma+s) \widetilde{\calJ}_j(\x_0,s)=\gamma \int_{\partial \calU_j} \widetilde{J}_j(\x,s|\x_0) d\x.
\end{equation}
In particular, for all $\x_0 \notin \calU_k$
\begin{equation}
\label{piRobin}
\pi_k(\x_0)\equiv \widetilde{\calJ}_k(\x_0,0)=\int_{\partial \calU_k} \widetilde{J}_k(\x,s|\x_0) d\x
\end{equation}
and 
\begin{align}
\label{TRobin}
T_k(\x_0)\equiv -\frac{1}{\pi_k(\x_0)}\left . \frac{\partial}{\partial s}\widetilde{\calJ}_k(\x_0,s)\right |_{s=0}= \frac{1}{\gamma} -\frac{1}{\pi_k(\x_0)}\int_{\partial \calU_k} \partial_s\widetilde{J}_k(\x,0|\x_0) d\x .
\end{align}

\begin{figure}[t!]
\centering
\includegraphics[width=12cm]{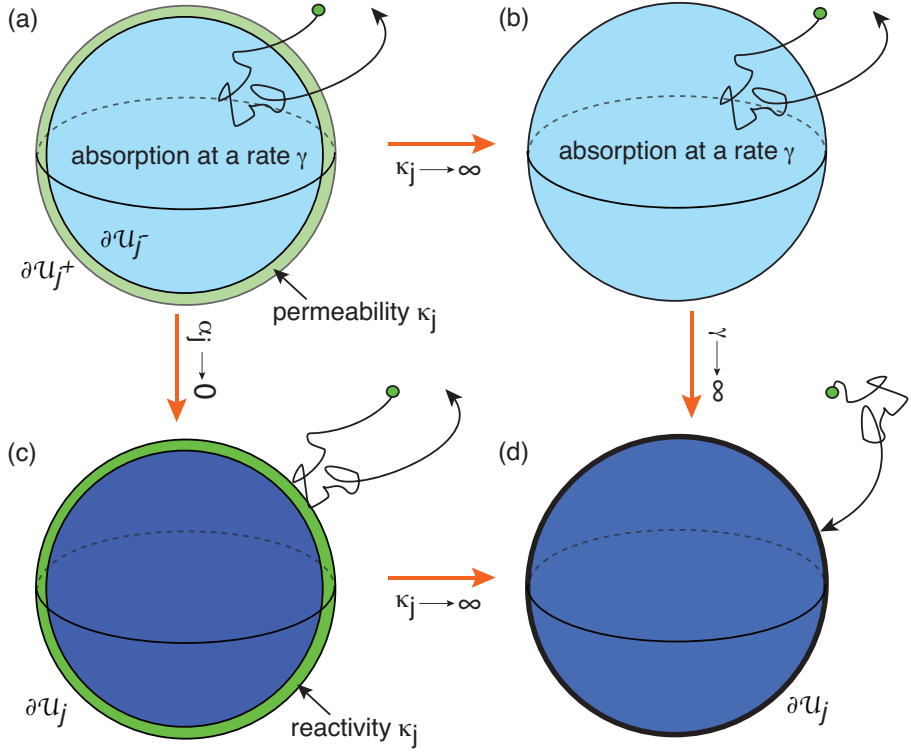} 
\caption{Limiting cases of the general semipermeable trap. (a) Semi-permeable interface $\partial \calU_j^{\pm}$ with a partially absorbing interior. (b) Totally permeable interface with a partially absorbing interior. (c) Partially absorbing surface $\partial \calU_j$. (d) Totally absorbing surface $\partial \calU_j$.}
\label{fig2}
\end{figure}

\subsection{Limiting cases}

The general model given by equations (\ref{master}) includes a number of limiting cases that have been considered previously. These are illustrated in Fig. \ref{fig2}. First, in the limit $\kappa_j\rightarrow \infty$ (complete permeability), the particle can freely enter and exit the $j$th trap and is absorbed at a rate $\gamma$ whenever it is within the trap interior (Fig. \ref{fig2}(b)). The corresponding narrow capture problem was analyzed in Ref. \cite{Bressloff22}. On the other hand, if $\alpha_j \rightarrow 0$, then once the particle enters the trap it cannot escape (zero outflux). This means that $\partial \calU_j$ acts as a partially absorbing boundary with reactivity $\kappa_j$, and the boundary condition (\ref{master}(c)) becomes Robin (Fig. \ref{fig2}(c)). That is, we have
\begin{subequations} 
\label{masterR}
\begin{align}
	\frac{\partial p(\x,t|\x_0)}{\partial t} &= D\nabla^2 p(\x,t|\x_0), \ \x\in \Omega\backslash \calU_a,	\ \nabla q\cdot \n=0,\ \x \in \partial \Omega .
	\end{align}
together with the Robin boundary conditions
\begin{align}
 D\nabla p(\x,t|\x_0)\cdot \n_j =\kappa_j  p(\x,t|\x_0) ,\quad \x \in \partial \calU_j
\end{align}
\end{subequations}
The narrow capture problem with partially reflecting surfaces $\partial \calU_j$ was recently developed with in the context of encounter-based models \cite{Bressloff22a}, see also section 7. Finally, in the double limit $\gamma\rightarrow \infty$ (or $\alpha_j\rightarrow 0$) and $\kappa_j\rightarrow \infty$, the surface $\calU_j$ becomes totally absorbing (Fig. \ref{fig2}(d)), and we recover the original example considered in Refs. \cite{Coombs15,Bressloff21B}.

\subsection{Backward equations, splitting probabilities and MFPTs}

In order to derive BVPs for the splitting probabilities and the MFPT we need to determine the backward diffusion equation. Since the adjoints of the semipermeable boundary conditions (\ref{master}c,d) are not as familiar as classical boundary conditions, we derive the backward equation explicitly.
We use the fact that the solution $p(\x,t|\x_0)$ to the forward BVP (\ref{master}) satisfies the Chapman-Kolmogorov equation 
\begin{equation}
\label{ChapK}
p(\x,t|\x_0)=\int_{\Omega} p(\x,t-\tau|\y)p(\y,\tau|x_0)d\y, \quad 0\leq \tau <\leq t.
\end{equation}
This essentially states that the probability of $\x$ given $\x_0$ is the sum of the probabilities of each possible path from $\x_0$ to $\x$. We have also imposed time translation invariance.
Using the fact that the left-hand side of equation (\ref{ChapK}) is independent of the intermediate time $\tau$, we have
\begin{align}
0&=\int_{\Omega}\partial_{\tau}p(\x,t-\tau|\y)p(\y,\tau|\x_0)d\y +\int_{\Omega}p(\x,t-\tau|\y)\partial_{\tau}p(\y,\tau|\x_0)d\y .
\end{align}
Since $p(\y,\tau|\x_0)$ satisfies the forward BVP equation, $\partial_{\tau}[p(\y,\tau |\x_0)]$ can be replaced by terms involving derivatives 
with respect to $\y$:
\begin{align*}
&\int_{\Omega}p(\x,t-\tau|\y)\partial_{\tau}p(\y,\tau|\x_0)d\y=D\int_{\Omega\backslash \calU_a} u(\x,t-\tau|\y)\nabla^2_{\y}p(\y,\tau|\x_0)d\y\\
&\quad +\sum_{k=1}^N \int_{ \calU_k} v_k(\x,t-\tau|\y)[D_k\nabla^2_{\y}q_k(\y,\tau|\x_0)-\gamma q_k(\y,\tau|\x_0)]d\y.
\end{align*}
(Here we are using the notation that for all $\x \in \Omega$, $p(\x,t-\tau|\y)=u(\x,t-\tau|\y)$ if $\y \in \Omega\backslash \calU_a$ and $p(\x,t-\tau|\y)=v_k(\x,t-\tau|\y)$  if $\y \in  \calU_k$.)
Integrating by parts using Green's identities shows that
\begin{align*}
&\int_{\Omega\backslash \calU_a} u(\x,t-\tau|\y)\nabla^2_{\y}p(\y,\tau|\x_0)d\y\\
&=- \int_{\Omega\backslash \calU_a} \nabla_{\y} u(\x,t-\tau|\y)\cdot \nabla_{\y}p(\y,\tau|\x_0) d\y+ \int_{\partial \Omega} u(\x,t-\tau|\y)\nabla_{\y}p(\y,\tau|\x_0)\cdot \n \, d\y\\
&- \sum_{k=1}^N \int_{\partial \calU_k^+} u(\x,t-\tau|\y) \nabla_{\y} p(\y,\tau|\x_0) \cdot \n_k d\y\\
&=\int_{\Omega\backslash \calU_a} \nabla_{\y}^2 u(\x,t-\tau|\y) p(\y,\tau|\x_0) d\y- \int_{\partial \Omega} p(\y,\tau|\x_0) \nabla_{\y} u(\x,t-\tau|\y)\cdot \n d\y\\
&\quad - \sum_{k=1}^N \int_{\partial \calU_k^+} u(\x,t-\tau|\y) \nabla_{\y} p(\y,\tau|\x_0) \cdot \n_k d\y \\
&\quad + \sum_{k=1}^N \int_{\partial \calU_k^+}  p(\y,\tau|\x_0) \nabla_{\y} u(\x,t-\tau|\y)\cdot \n_kd\y.
\end{align*}
Similarly,
\begin{align*}
\int_{ \calU_k} v_k(\x,t-\tau|\y) \nabla^2_{\y}q_k(\y,\tau|\x_0)&=\int_{\calU_k} \nabla_{\y}^2 v_k(\x,t-\tau|\y) p_k(\y,\tau|\x_0) d\y\\
&\quad +\int_{\partial \calU_k^-} v_k(\x,t-\tau|\y^-) \nabla_{\y} q_k(\y,\tau|\x_0) \cdot \n_k d\y \\
&\quad -  \int_{\partial \calU_k^-}  q_k(\y,\tau|\x_0) \nabla_{\y} v_k(\x,t-\tau|\y)\cdot \n_kd\y.
\end{align*}

We see from the above equations that the various boundary terms cancel if we impose the adjoint boundary conditions $\nabla_{\y}u(\x,t-\tau|\y)\cdot \n=0$ and
\begin{align*}
&D p(\y^+,\tau|\x_0) \nabla_{\y} u(\x,t-\tau|\y^+)\cdot \n_k-Du(\x,t-\tau|\y^+) \nabla_{\y} p(\y^+,\tau|\x_0) \cdot \n_k \\
&=D_kq_k(\y^-,\tau|\x_0) \nabla_{\y} v_k(\x,t-\tau|\y^-)\cdot \n_k-D_kv_k(\x,t-\tau|\y^-) \nabla_{\y} q_k(\y^-,\tau|\x_0) \cdot \n_k
\end{align*}
for all $\y^{\pm}\in \partial \calU_k^{\pm}$. Setting 
\begin{equation}
J_k^{\dagger}(\x,t|\y^+) = D  \nabla_{\y} u(\x,t|\y^+)\cdot \n_k,\quad J_k^{\dagger}(\x,t|\y^-) =D_k  \nabla_{\y} v_k(\x,t-\tau|\y^-)\cdot \n_k
\end{equation}
gives
\begin{align*}
& p(\y^+,\tau|\x_0) J_k^{\dagger}(\x,t-\tau|\y^+) -u(\x,t-\tau|\y^+) J_k(\y,\tau|\x_0)  \\
&=  q_k(\y^-,\tau|\x_0) J_k^{\dagger}(\x,t-\tau|\y^-) - v_k(\x,t-\tau|\y^-) J_k(\y,\tau|\x_0) ,\ \y^{\pm}\in \partial \calU_k^{\pm}.
\end{align*}
Rearranging the last equation and imposing the forward semipermeable boundary conditions (\ref{master}c,d) yields
\begin{align}
& p(\y^+,\tau|\x_0) J_k^{\dagger}(\x,t-\tau|\y^+) - q_k(\y^-,\tau|\x_0) J_k^{\dagger}(\x,t-\tau|\y^-) \nonumber \\
&=  [u(\x,t-\tau|\y^+) - v_k(\x,t-\tau|\y^-)] J_k(\y,\tau|\x_0) \\
&=[u(\x,t-\tau|\y^+) - v_k(\x,t-\tau|\y^-)] \kappa_k[(1-\alpha_k)p(\y^+,\tau|\x_0)-\alpha_kq_k(\y^-,\tau|\x_0)]\nonumber 
\end{align}
for $\y^{\pm}\in \partial \calU_k^{\pm}$. We thus have the adjoint equations
\begin{subequations}
\begin{align}
J_k^{\dagger}(\x,t|\y^+) &= \kappa_k(1-\alpha_k) [u(\x,t|\y^+)-v_k(\x,t|\y^-)]\\
J_k^{\dagger}(\x,t|\y^-) &= \kappa_k \alpha_k[u(\x,t|\y^+)-v_k(\x,t|\y^-)] .
\end{align}
\end{subequations}
Combining our various results we  the following backward BVP
\begin{subequations} 
\label{masterb}
\begin{align}
	\frac{\partial u(\x,t|\y)}{\partial t} &= D\nabla_{\y}^2 u(\x,t|\y), \ \y\in \Omega\backslash \calU_a,\\
	\frac{\partial v_j(\x,t|\y)}{\partial t} &= D_j\nabla_{\y}^2 v_j(\x,t|\y) -\gamma v_j(\x,t|\y),\ \y\in \calU_j,
	\end{align}
	 with the adjoint semipermeable boundary conditions
\begin{align}
\alpha_jD\nabla_{\y} u(\x,t|\y^+)\cdot \n_j  &= (1-\alpha_j) D_j\nabla v_j(\x,t|\y^-)\cdot \n_j\equiv\frac{1}{2} J_j^{\dagger}(\x,t|\y) \\
J_j^{\dagger}(\x,t|\y) &= 2\kappa_j \alpha_j(1-\alpha_j) [u(\x,t|\y^+)-v_j(\x,t|\y^-)] ,\ \y\in \partial \calU,
\end{align}
and the exterior boundary condition
\begin{equation}
\nabla_{\y} u(\x,t|\y)\cdot \n=0,\ \y \in \partial \Omega .
\end{equation}
\end{subequations}
Note that if $\alpha_j=1/2$ (unbiased interface), then the semipermeable boundary conditions are self-adjoint.

We can now write down the corresponding BVP for the absorption flux $\calJ_k(\x_0,t)$ by integrating equations (\ref{masterb}) with respect to $\x\in \calU_k$:
\begin{subequations} 
\label{masterQ}
\begin{align}
	\frac{\partial \calJ_k(\x_0,t)}{\partial t} &= D\nabla^2 \calJ_k(\x_0,t), \ \x_0 \in \Omega\backslash \calU_a,\\
	\frac{\partial \calJ_k(\x_0,t)}{\partial t} &= D_j\nabla^2 \calJ_k(\x_0,t) -\gamma \calJ_k(\x_0,t),\ \x_0\in \calU_j,
	\end{align}
together with the adjoint semipermeable boundary conditions
\begin{align}
\alpha_jD\nabla \calJ_k(\x_0,t)\cdot \n_j  &= (1-\alpha_j) D_j\nabla \calJ_k(\x_0,t)\cdot \n_j\equiv\frac{1}{2} J_{jk}^{\dagger}(\x_0,t) \\
J_{jk}^{\dagger}(\x_0,t) &= 2\kappa_j \alpha_j(1-\alpha_j) [\calJ_k(\x_0^+,t)-\calJ_k(\x_0^-,t)] ,\ \x_0\in \partial \calU_j,
\end{align}
and the exterior boundary condition
\begin{equation}
\nabla  \calJ_k(\x_0,t)\cdot \n=0,\ \x_0\in \partial \Omega .
\end{equation}
\end{subequations}
All derivatives are with respect to the initial position $\x_0$. The initial condition is
\begin{equation}
\calJ_k(\x_0,0)=\gamma {\mathbb I}_{\calU_k}(\x_0),
\end{equation}
where ${\mathbb I}_{\calU_k}(\x_0)=1$ if $\x_0\in \calU_k$ and is zero otherwise.
Laplace transforming the BVP (\ref{masterQ}) and using equation (\ref{split}), we find that the splitting probability $\pi_k(\x)$ satisfies the BVP
\begin{subequations} 
\label{BVPsplit}
\begin{align}
& D\nabla^2 \pi_k(\x_0) = 0, \ \x_0 \in \Omega\backslash \calU_a,\\
 &D_j\nabla^2 \pi_k(\x_0) +\gamma (\delta_{j,k}-\pi_k(\x_0))=0,\ \x_0\in \calU_j,\\
\alpha_jD\nabla \pi_k(\x_0^+)\cdot \n_j  &= (1-\alpha_j) D_j\nabla \pi_k(\x_0^-)\cdot \n_j\nonumber  \\
&= \kappa_j \alpha_j(1-\alpha_j) [\pi_k(\x_0^+)-\pi_k(\x_0^-)] ,\ \x_0\in \partial \calU_j,
\end{align}
and the exterior boundary condition
\begin{equation}
\nabla  \pi_k(\x_0)\cdot \n=0,\ \x_0\in \partial \Omega .
\end{equation}
\end{subequations}
Similarly, using equation (\ref{Tuncon}), we obtain the BVP for the unconditional MFPT, which takes the form
\begin{subequations}
\label{BVPT}
\begin{align}
\nabla^2 T(\x_0)&=-\frac{1}{D}, \  \x\in \Omega\backslash \calU_a, \\
\nabla^2 T(\x_0)-\gamma T(\x_0)&=-\frac{1}{D_j}, \  \x\in \calU_j, \\
\alpha_jD\nabla T(\x_0)\cdot \n_j  &= (1-\alpha_j) D_j\nabla  T(\x_0) \cdot \n_j\nonumber \\
&= \kappa_j \alpha_j(1-\alpha_j) [T(\x_0^+)-T(\x_0)^-] ,\ \x_0\in \partial \calU_j,
\end{align}
and the exterior boundary condition
\begin{equation}
\nabla  T(\x_0)\cdot \n=0,\ \x_0\in \partial \Omega .
\end{equation}
\end{subequations}

\section{Matched asymptotic analysis: splitting probabilities}

The BVP (\ref{BVPsplit}) for the splitting probabilities and the BVP (\ref{BVPT}) for the unconditional MFPT can both be solved in the small trap limit using matched asymptotic expansions and Green's functions along analogous lines to previous studies of the narrow capture problem \cite{Cheviakov11,Chevalier11,Coombs15,Bressloff15,Lindsay15,Lindsay17,Bressloff21B,Bressloff22}. The basic idea is to construct an inner or local solution valid in an $O(\epsilon)$ neighborhood of each trap, and then match to an outer or global solution that is valid away from each neighborhood. We first consider the BVP for the splitting probabilities.

In the outer region $\Omega'=\Omega\backslash \{\x_1,\ldots,\x_N\}$, $\pi_k$ is expanded as
$\pi_k=\pi_k^{(0)}+ \epsilon \pi_k^{(1)}+\epsilon^2 \pi_k^{(2)}+\ldots$, where $\pi_k^{(0)}$ is an unknown constant.  After dropping the subscript on $\x_0$, we have
\begin{subequations} 
\label{outersplit}
\begin{align}
 D\nabla^2 \pi_k^{(n)}(\x) &= 0, \ \x \in \Omega',\\
\nabla  \pi_k^{(n)}(\x)\cdot \n&=0,\ \x\in \partial \Omega ,\quad n \geq 1,
\end{align}
\end{subequations}
together with a set of singularity conditions as $\x\rightarrow \x_j$, $j=1,\ldots,N$.
The latter are determined by matching to the inner solution. In the inner region around the $j$-th trap, we introduce the stretched coordinates ${\bf z}=\epsilon^{-1}(\x-\x_j)$ and set 
\[U_{k}({\bf z}) =\pi_k(\x_j+\epsilon \z),\ |\z| < \rho_j,\quad  V_{k}({\bf z}) =\pi_k(\x_j+\epsilon \z),\ |\z|> \rho_j.\]
Note that $U_k$ and $V_k$ are the components of the inner solution corresponding to the interior and exterior of the $j$th trap. For the moment, we will assume that $\gamma=O(1/\epsilon^2  )$ and $\kappa=O(1/\epsilon)$ and perform the scalings
  \begin{equation}
  \label{scale1}
  \gamma=\frac{\gamma'}{\epsilon^2},\quad  \kappa_j= \frac{\kappa'_j}{\epsilon} .
  \end{equation} 
 Introducing the series expansions \begin{equation}
 U_{k}=U_k^{(0)}+\epsilon U_{k}^{(1)}+\ldots,\quad V_k=V_{k}^{(0)}+\epsilon V_{k}^{(1)}+\ldots,
 \end{equation}
we have
\begin{subequations} 
\label{innersplit}
\begin{align}
&D\nabla^2  V_k^{(n)}(\z)=0,\ |\z|>\rho_j,\\
& D_j\nabla^2 U_k^{(n)}(\z) + \gamma' (\delta_{j,k}\delta_{n,0}-U_k^{(n)}(\z))=0,\ |\z|<\rho_j,\\
\alpha_jD\nabla V_k^{(n)}(\z) \cdot \n_j  &= (1-\alpha_j) D_j\nabla U_k^{(n)}(\z) \cdot \n_j\nonumber \\
& = \kappa_j' \alpha_j(1-\alpha_j) [U_k^{(n)}(\z)-V_k^{(n)}(\z)] ,\ |\z|=\rho_j,
\end{align}
\end{subequations}
Finally, the matching condition is that the near-field behavior of the outer solution as $\x\rightarrow \x_j$ should agree with the far-field behavior of the inner solution as $|\y|\rightarrow \infty$, which is expressed as 
\[ \pi_k^{(0)} +\epsilon \pi_k^{(1)} +\epsilon^2 \pi_k^{(2)}\sim V_k^{(0)} +\epsilon V_k^{(1)}+\epsilon^2 V_k^{(2)}\ldots.
\]

\subsection{Inner solution $U_k^{(0)},V_k^{(0)}$} We solve the inner BVP for $n=0$ using spherical polar coordinates with $\rho=|\z|$ and the far-field condition $V_k^{(0)}\sim \pi_k^{(0)} = {\rm constant}$. Equations (\ref{innersplit}) reduce to the form  
\begin{subequations}
\label{sph}
\begin{align}
 &D\frac{\partial^2V_k^{(0)}(\rho)}{\partial \rho^2} + \frac{2D}{\rho}\frac{\partial V_k^{(0)}(\rho)}{\partial \rho}=0, \ \rho>\rho_j,\\
&D_j\frac{\partial^2U_k^{(0)}(\rho)}{\partial \rho^2} + \frac{2D_j}{\rho}\frac{\partial U_k^{(0)}(\rho)}{\partial \rho}+\gamma' (\delta_{j,k}-U_k^{(0)}(\rho) )=0 ,\  \rho<\rho_j,\\
  &\alpha_jD\frac{\partial V_k^{(0)}(\rho_j)}{\partial \rho}=(1-\alpha_j)D_j\frac{\partial U_k^{(0)}(\rho_j)}{\partial \rho}=\kappa_j'\alpha_j(1-\alpha_j)[V_k^{(0)}(\rho_j) -U_k^{(0)}(\rho_j)].\end{align}
  \end{subequations}
A well known trick for solving the spherically symmetric diffusion equation in 3D is to perform the change of variables $U_k^{(0)}(\rho)=\phi_k^{(0)}(\rho)/\rho$ and $V_k^{(0)}(\rho)=\psi_k^{(0)}(\rho)/\rho$. Substituting into equations (\ref{sph}a,b) yields 
\begin{subequations}
\label{sph2}
\begin{align}
 &D\frac{\partial^2\psi_k^{(0)}(\rho)}{\partial \rho^2} =0,\   \rho >\rho_j ,\\
 &D_j\frac{\partial^2\phi_k^{(0)}(\rho)}{\partial \rho^2} +\gamma' (\rho\delta_{j,k}-\phi_k^{(0)}(\rho))=0 ,\  \rho <\rho_j
\end{align}
\end{subequations}
We also have the far-field condition $\psi_k^{(0)} \sim \pi_k^{(0)} \rho$. We thus obtain the general solution
\begin{equation}
\label{1Dsola}
U_k^{(0)}(\rho)=  \delta_{j,k}+B_{jk}^{(0)}\frac{ \sinh(\sqrt{\gamma'/D_j}\rho)}{\rho} ,\quad V_k^{(0)}(\rho)=\frac{A_{jk}^{(0)}}{\rho}+\pi_k^{(0)} .
\end{equation}

Substituting into the boundary conditions (\ref{sph}c) gives
\begin{subequations}
\begin{align}
-\frac{\alpha_jDA_{jk}^{(0)}}{\rho_j}&=(1-\alpha_j)B_{jk}^{(0)}\left \{\sqrt{\gamma' D_j}\cosh(\sqrt{\gamma'/D_j}\rho_j) - \frac{D_j}{\rho_j}\sinh(\sqrt{\gamma'/D_j}\rho_j) \right \},\\
-\frac{\alpha_jDA_{jk}^{(0)}}{\rho_j}&=\kappa_j'\alpha_j(1-\alpha_j)\left [A_{jk}^{(0)}+[\pi_k^{(0)}-\delta_{j,k}] \rho_j  - B_{jk}^{(0)}\sinh(\sqrt{\gamma'/D_j}\rho_j) \right ].
\end{align}
\end{subequations}
 Setting
 \begin{equation}
\Gamma^a_{j}=\frac{\alpha_jD}{(1-\alpha_j)\rho_j},\quad\Gamma^b_{j}=\frac{D_j}{\rho_j}\left [\frac{D}{(1-\alpha_j)\kappa_j'\rho_j}  +1\right ]
\end{equation}
and
\begin{equation}
\label{Fj}
\calC_j=\frac{D_j\left \{\rho_j\sqrt{\gamma' /D_j}\cosh(\sqrt{\gamma'/D_j}\rho_j) -  \sinh(\sqrt{\gamma'/D_j}\rho_j)\right \}}{\Gamma_j^a \sinh(\sqrt{\gamma'/D_j}\rho_j)+\Gamma_j^b \left \{\rho_j\sqrt{\gamma' /D_j}\cosh(\sqrt{\gamma'/D_j}\rho_j)-\sinh(\sqrt{\gamma'/D_j}\rho_j)\right \}},
\end{equation}
we find that
\begin{align}
\label{Aj}
A_{jk}^{(0)} &=\calC_j(\delta_{j,k}-\pi_k^{(0)}).
\end{align}
The expression for the coefficient $\calC_j$ is one of the main results of this paper. Its significance will be developed below.

\subsection{Calculation of $\pi_k^{(0)}$}

So far we have shown that the inner solution $V_k^{(0)}$ outside the $j$th trap is
\begin{equation}
V_k^{(0)}(\z) = \pi_k^{(0)}+\frac{A_{jk}^{(0)}}{|\z|},
\end{equation}
with $A_{jk}^{(0)}$ given by equation (\ref{Aj}). It follows that the outer solution $\pi_k^{(1)} $ satisfies the singularity condition
\[\pi_k^{(1)} (\x)\sim \frac{A_{jk}^{(0)}}{|\x-\x_j|}\quad \mbox{as } \x\rightarrow \x_j.\]
In other words, $\pi_k^{(1)} $ satisfies the inhomogeneous equation
\begin{align}
\pcb{\nabla^2 \pi_k^{(1)}(\x) =-4\pi \sum_{j=1}^NA_{jk}^{(0)}\delta(\x-\x_j), \ \x \in\Omega; \quad \nabla \pi_k^{(1)} (\x)\cdot \n=0,\quad \x \in \partial \Omega.}
 \label{asym3D}
\end{align}
Integrating this equation with respect to $\x\in \Omega$ and using the divergence theorem implies that
$\sum_{j=1}^NA_{jk}^{(0)}=0$.
Hence, summing equation (\ref{Aj}) with respect to $j=1,\ldots,N$ determines the unknown constant $\pi_k^{(0)}$ according to
\begin{equation}
\label{pi0}
\pi_k^{(0)} =\frac{\calC_k}{N\overline{\calC}} ,\quad \overline{\calC}=\frac{1}{N}\sum_{j=1}^N \calC_j.
\end{equation}
In particular, note that $\sum_{k=1}^N\pi_k^{(0)}=1$ and thus $\sum_{k=1}^N\pi_k^{(n)}=0$ for $n\geq 1$.

Equation (\ref{asym3D}) can now be solved in terms of the Neumann Green's function:
\begin{equation}
\label{pi1}
\pi_k^{(1)}(\x)=4\pi \sum_{j=1}^NA_{jk}^{(0)}G(\x,\x_j)+\chi_{k}^{(1)},
\end{equation}
where
\begin{subequations}
\begin{align}
\nabla^2 G(\x,\x')&=\frac{1}{|\Omega|} -\delta(\x-\x'),\, \pcb{\x\in \Omega};\ \nabla G\cdot \n=0,\, \pcb{\x \in \partial \Omega},\\
G(\x,\x')&=\frac{1}{4\pi|\x-\x'|}+R(\x,\x'),\ \int_{\Omega}G(\x,\x')d\x=0,
\end{align}
\end{subequations}
with $R(\x,\x')$ corresponding to the regular part of the Green's function. Integrating both sides of equation (\ref{pi1}) with respect to $\x \in \Omega$ shows that the unknown constant $\chi_k^{(1)}$ satisfies
\begin{equation}
\chi_{k}^{(1)}=\frac{1}{|\Omega|}\int_{\Omega}\pi_k^{(1)}(\x)d\x,\quad \sum_{k=1}^N \chi_k^{(1)}=0.
\end{equation}

\subsection{Calculation of $\chi_{k}^{(1)}$} The outer solution (\ref{pi1}) implies that the inner solution $V_k^{(1)}$ around the $j$th trap has the far-field behavior
\begin{equation}
\label{lam1}
V_k^{(1)}(\z)\sim \Lambda_{jk}^{(1)} \equiv 4\pi \sum_{i=1}^NG_{ji}A_{ik}^{(0)}+\chi_{k}^{(1)},\quad |\z|\rightarrow \infty
\end{equation}
where 
\begin{equation}
G_{ji}=G(\x_j,\x_i) \mbox{ for } i\neq j,\quad G_{jj}=R(\x_j,\x_j).
\end{equation}
The inner BVP for $n=1$ for $V_k^{(1)}(\rho)=\psi_k^{(1)}(\rho)/\rho$ and $U_k^{(1)}(\rho)=\phi_k^{(1)}(\rho)/\rho$ is
\begin{subequations}
\label{1sph2}
\begin{align}
 D\frac{\partial^2\psi_k^{(1)}(\rho)}{\partial \rho^2} &=0,\   \rho >\rho_j ,\\
 D_j\frac{\partial^2\phi_k^{(1)}(\rho)}{\partial \rho^2} -\gamma' \phi_k^{(1)}(\rho)&=0 ,\  \rho <\rho_j,
\end{align}
\end{subequations}
and $\psi_k^{(1)} \sim \Lambda_{jk}^{(1)} \rho$. The general solution is thus
\begin{equation}
\label{11Dsola}
U_k^{(1)}(\rho)= B_{jk}^{(1)}\frac{\sinh(\sqrt{\gamma'/D_j}\rho)}{\rho} ,\quad V_k^{(1)}(\rho)=\frac{A_{jk}^{(1)}}{\rho}+\Lambda_{jk}^{(1)} .
\end{equation}
Proceeding along analogous lines to the case $n=0$, we find that
\begin{align}
\label{1Aj}
A_{jk}^{(1)} &=-\calC_j \Lambda_{jk}^{(1)},
\end{align}
with $\calC_j$ defined in equation (\ref{Fj}). Hence, the inner solution $V_k^{(1)}$ outside the $j$th trap is
\begin{equation}
V_k^{(1)}(\z)=\left (1-\frac{\calC_j}{\z}\right )\Lambda_{jk}^{(1)}
\end{equation}

It follows that the outer solution $\pi_k^{(2)}$ has the singular behavior
\begin{equation}
\pi_k^{(2)}\sim -\frac{\calC_j\Lambda_{jk}^{(1)}}{|\x-\x_j|} \mbox{ as } \x\rightarrow \x_j.
\end{equation}
That is, $\pi_k^{(2)} $ satisfies the inhomogeneous equation
\begin{align}
\pcb{\nabla^2 \pi_k^{(2)}(\x) =4\pi \sum_{j=1}^N\calC_j\Lambda_{jk}^{(1)}\delta(\x-\x_j), \ \x \in\Omega; \quad \nabla \pi_k^{(2)} (\x)\cdot \n=0,\quad \x \in \partial \Omega.}
\end{align}
Integrating this equation with respect to $\x\in \Omega$ and using the divergence theorem implies that
\begin{equation}
\sum_{j=1}^N\calC_j \left (4\pi \sum_{i=1}^NG_{ji}[\calC_i(\delta_{i,k}-\pi_k^{(0)})]+\chi_{k}^{(1)}\right )=0,
\end{equation}
where we have used equations (\ref{Aj}) and (\ref{lam1}). That is,
\begin{align}
\chi_k^{(1)}=\frac{4\pi}{N\overline{\calC}} \left ( \pi_k^{(0)}\sum_{i,j=1}^N \calC_jG_{ji}\calC_i-\sum_{j=1}^N \calC_jG_{jk}\calC_k\right )
\end{align}

\begin{figure}[b!]
\centering
\includegraphics[width=13cm]{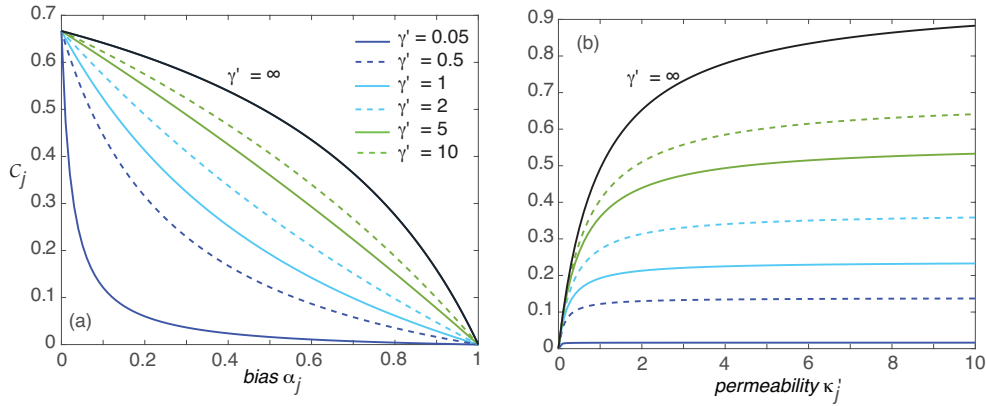} 
\caption{Plot of coefficient $\calC_j$ defined by equation (\ref{Fj}) as a function of (a) the bias parameter $\alpha_j$ and (b) the permeability $\kappa_j'$ for different absorption rates $\gamma'$. Default parameters are $D=D_j=\rho_j=1=\kappa_j'=1$ and $\alpha_j=0.5$.}
\label{fig3}
\end{figure}

\subsection{Properties of $\calC_j$}
In summary, the outer solution for the splitting probability $\pi_k$ is 
\begin{align}
\pi_k(\x_0)&\sim \frac{\calC_k}{N\overline{\calC}} \bigg \{1+ 4\pi \epsilon \left (\sum_{i,j=1}^N\frac{1}{N\overline{\calC} } \calC_jG_{ji}\calC_i-\sum_{j=1}^N \calC_jG_{jk}\right ) \nonumber \\
&\hspace{3cm}+4\pi \epsilon \sum_{j=1}^N\calC_j[G(\x_0,\x_k)-G(\x_0,\x_j)]\bigg \}+O(\epsilon^2).
\label{bigp}
\end{align}
Hence, the dependence of $\pi_k(\x_0)$ on the properties of the $j$th trap ($\kappa_j,\alpha_j,D_j,\rho_j)$ are encapsulated by the coefficients $\calC_j$ in equation (\ref{Fj}). In the limit $\gamma' \rightarrow \infty$, a particle is instantaneously absorbed when entering a trap and 
\begin{equation}
\label{Finf}
\lim_{\gamma'\rightarrow \infty} \calC_j=\frac{D_j}{\Gamma^b_{j}}={\rho_j}\left [\frac{D}{(1-\alpha_j)\kappa_j'\rho_j}  +1\right ]^{-1}.
\end{equation}
If we also take $\kappa_j'\rightarrow \infty$, $j=1,\ldots,N$, then the interfaces become totally permeable and $\calC_j\rightarrow \rho_j$. The corresponding splitting probabilities reduce to the form
\begin{align}
\pi_k(\x_0)&\sim \frac{\rho_k}{N\overline{\rho}} \bigg \{1+ 4\pi \epsilon \left (\sum_{i,j=1}^N\frac{1}{N\overline{\rho} } \rho_jG_{ji}\rho_i-\sum_{j=1}^N \rho_jG_{jk}\right ) \nonumber \\
&\hspace{3cm}+4\pi \epsilon \sum_{j=1}^N\rho_j[G(\x_0,\x_k)-G(\x_0,\x_j)]\bigg \}+O(\epsilon^2).
\end{align}
This is identical to the result previously derived for the 3D narrow capture problem with totally absorbing surfaces $\partial \calU_j$ \cite{Coombs15,Bressloff21B}. The length $\rho_j$ is the diffusive capacitance of a sphere of radius $\rho_j$. 

Now suppose that $\gamma'$ is finite, $\alpha_i=0.5$ and $\kappa_j'\rightarrow \infty$, see Fig.\ref{fig2}(b). We then recover the example of partially absorbing target interiors analyzed in Ref. \cite{Bressloff22}, except that here we are allowing the diffusivity to be discontinuous across each completely permeable interface. From equation (\ref{Fj}),
\begin{equation}
\lim_{\kappa_j' \rightarrow \infty} \calC_j= \frac{\rho_jD_j\left \{\rho_j\sqrt{\gamma' /D_j}\cosh(\sqrt{\gamma'/D_j}\rho_j) -  \sinh(\sqrt{\gamma'/D_j}\rho_j)\right \}}{[D-D_j]\sinh(\sqrt{\gamma'/D_j}\rho_j)+D_j  \rho_j\sqrt{\gamma' /D_j}\cosh(\sqrt{\gamma'/D_j}\rho_j) },
\end{equation}
In particular, if $D_j=D$ then
\begin{equation}
\label{Finf2}
\lim_{\kappa_j' \rightarrow \infty} \calC_j=\rho_j \left \{1-  \frac{\tanh(\sqrt{\gamma'/D}\rho_j)}{   \rho_j\sqrt{\gamma' /D} }\right \},
\end{equation}
which recovers our previous result \cite{Bressloff22}. 
In the case of finite $\kappa'_j$ and $\gamma'$, the general expression for $\calC_j$ in equation (\ref{Fj}) can be interpreted as the effective capacitance of the sphere with $\calC_j\leq \rho_j$. Reducing the permeability $\kappa_j'$ or increasing the outward bias $\alpha_j$ reduces $\calC_j$, since it is more difficult for a particle to enter the trap and be absorbed. The latter is illustrated in Fig. \ref{fig3}(a), which shows plots of $\calC_j$ against $\alpha_j$ for various absorption rates $\gamma'$. In the limit $\gamma' \rightarrow \infty$ we that the range of $\calC_j$ is $[0,2/3]$, which is consistent with equation (\ref{Finf}) for $D=D_j=\rho_j=\kappa_j'=1$. Corresponding plots of $\calC_j$ as a function of $\kappa_j'$ are shown in Fig. \ref{fig3}(b).

Finally, consider the intermediate absorption regime in which $\epsilon \ll \gamma' \ll 1$ and $D_j/\rho_j^2=O(1) $ for all $ j=1,\ldots,N$. Taylor expanding the numerator and denominator on the right-hand side of equation (\ref{Fj}) with respect to $\xi_j=\sqrt{\gamma'/D_j}\rho_j$ gives 
\begin{align}
\calC_j&=\frac{D_j\left \{\xi_j(1+\xi_j^2/2)) -  \xi_j(1+\xi_j^2/6) \right \}+\ldots }{\Gamma_j^a\xi_j( 1+\xi_j^2/6)+\Gamma_j^b \left \{\xi_j(1+\xi_j^2/2)) -  \xi_j(1+\xi_j^2/6)  \right \}+\ldots}\nonumber \\
&\approx \frac{D_j\xi_j^2}{3\Gamma_j^a }= \frac{\gamma'(1-\alpha_j)\rho_j^3}{3\alpha_jD}=\frac{\gamma'(1-\alpha_j)|\calU^0_j|}{4\pi D\alpha_j}
\label{app1}
\end{align}
That is, to leading order, $\calC_j$ is proportional to the absorption rate $\gamma'$ and the stretched trap volume $|\calU^0_j|=4\pi \rho_j^3/3$, whereas it is independent of the permeability $\kappa_j'$. On the other hand, in the intermediate permeability regime $\epsilon \ll \kappa_j' \ll 1$ and $(1-\alpha_j)\rho_j/D=O(1)$ for all $ j=1,\ldots,N$, we find that
\begin{equation}
\calC_j\approx \frac{D_j}{\Gamma_j^b}\approx \frac{(1-\alpha_j)\kappa_j'\rho_j^2} {D} =\frac{(1-\alpha_j)\kappa_j'|\partial \calU^0_j|} {4\pi D} .
\label{app2}
\end{equation}
To leading order, $\calC_j$ is proportional to the permeability $\kappa_j'$ and the stretched trap surface area $|\partial \calU_j^0|=4\pi \rho_j^2$, but is independent of $\gamma'$.  

\section{Matched asymptotic analysis: MFPT}

A similar analysis to section 3 can be performed in the case of the MFPT. Again we will consider the particular scaling (\ref{scale1}).
In the outer region, $T$ is expanded as
$T={\epsilon}^{-1}T^{(0)}+ T^{(1)}+\epsilon T^{(2)}+\ldots$
Here $T_0$ is an unknown constant, and
\begin{align}
\nabla^2 T^{(n)}(\x)&=-\frac{\delta_{n,1}}{D},\, \x\in \Omega\backslash \{\x_1,\ldots,\x_N\},\quad \nabla T^{(n)}(\x)\cdot \n=0,\, \x\in \partial \Omega
\end{align}
 together with certain singularity conditions as $\x\rightarrow \x_j$, $j=1,\ldots,N$. The latter are determined by matching to the inner solution. In the inner region around the $j$-th trap, we again introduce the stretched coordinates ${\bf z}_j=\epsilon^{-1}(\x-\x_j)$ and set 
 \[U({\bf z}) =T(\x_j+\epsilon \z),\ |\z| < \rho_j,\quad  V({\bf z}) =T(\x_j+\epsilon \z),\ |\z|> \rho_j.\]
 We keep the same scalings $\gamma\rightarrow\gamma/\epsilon^2$ and $\kappa_j\rightarrow \kappa_j/\epsilon$. Introducing the series expansions
  \begin{equation}
 U=\epsilon^{-1}U^{(0)}+ U^{(1)}+\ldots,\quad V=\epsilon^{-1}V^{(0)}+ V^{(1)}+\ldots,
 \end{equation}
we have
\begin{subequations} 
\label{innerT}
\begin{align}
&D\nabla^2  V^{(n)}(\z)=0,\ |\z|>\rho_j,\\
& D_j\nabla^2 U^{(n)}(\z) - \gamma U^{(n)}(\z)=0,\ |\z|<\rho_j,\ n=0,1,\\
\alpha_jD\nabla V^{(n)}(\z) \cdot \n_j  &= (1-\alpha_j) D_j\nabla U^{(n)}(\z) \cdot \n_j\nonumber \\
& = \kappa_j \alpha_j(1-\alpha_j) [U^{(n)}(\z)-V^{(n)}(\z)] ,\ |\z|=\rho_j,
\end{align}
\end{subequations}
Finally, the matching condition is that the near-field behavior of the outer solution as $\x\rightarrow \x_j$ should agree with the far-field behavior of the inner solution as $|\y|\rightarrow \infty$, which is expressed as 
\[\epsilon^{-1} T^{(0)} + T^{(1)} +\epsilon T^{(2)}\sim \epsilon^{-1}V^{(0)} + V^{(1)}+\epsilon V^{(2)}\ldots.
\]

Following along similar lines to section 3.1 by working in spherical polar coordinates, we obtain the solutions
\begin{equation}
\label{Tsola}
U^{(0)}(\rho)=B_{j}^{(0)} \frac{\sinh(\sqrt{\gamma/D_j}\rho)}{\rho} ,\quad V^{(0)}(\rho)=\frac{A_{j}^{(0)}}{\rho}+T^{(0)}.
\end{equation}
with
\begin{subequations}
\begin{align}
-\frac{\alpha_jDA_{j}^{(0)}}{\rho_j}&=(1-\alpha_j)B_{j}^{(0)}\left \{\sqrt{\gamma D_j}\cosh(\sqrt{\gamma/D_j}\rho_j) - \frac{D_j}{\rho_j}\sinh(\sqrt{\gamma/D_j}\rho_j) \right \},\\
-\frac{\alpha_jDA_{j}^{(0)}}{\rho_j}&=\kappa_j\alpha_j(1-\alpha_j)\left [A_{j}^{(0)}+T^{(0)} \rho_j  - B_{j}^{(0)}\sinh(\sqrt{\gamma/D_j}\rho_j) \right ].
\end{align}
\end{subequations}
Hence,
$A_{j}^{(0)} =-\calC_jT^{(0)}$ with $\calC_j$ defined in equation (\ref{Fj}).
Next, the outer solution $T^{(1)} $ satisfies the singularity condition
\[T^{(1)} (\x)\sim -\frac{\calC_jT^{(0)}}{|\x-\x_j|}\quad \mbox{as } \x\rightarrow \x_j.\]
Hence, $T^{(1)} $ satisfies the inhomogeneous equation
\begin{align}
D\nabla^2 T^{(1)}(\x) &=-\frac{1}{D}+4\pi \sum_{j=1}^N\calC_jT^{(0)}\delta(\x-\x_j), \ \x \in\Omega, \ \nabla T^{(1)} (\x)\cdot \n=0,\quad \x \in \partial \Omega.
 \label{Tasym3D}
\end{align}
Integrating with respect to $\x\in \Omega$ and using the divergence theorem implies that
\begin{equation}
T^{(0)}=\frac{|\Omega|}{4\pi DN\overline{\calC}}.
\end{equation}

Equation (\ref{Tasym3D}) can now be solved in terms of the Neumann Green's function:
\begin{equation}
\label{Tpi1}
T^{(1)}(\x)=-4\pi T^{(0)}\sum_{j=1}^N\calC_{j}G(\x,\x_j)+\chi^{(1)},
\end{equation}
Integrating both sides of equation (\ref{Tpi1}) with respect to $\x \in \Omega$ implies that the unknown constant $\chi^{(1)} $ satisfies
\begin{equation}
\chi^{(1)}=\frac{1}{|\Omega|}\int_{\Omega}T^{(1)}(\x)d\x.
\end{equation}
The outer solution (\ref{Tpi1}) implies that the inner solution $V^{(1)}$ around the $j$th trap has the far-field behavior
\begin{equation}
\label{Tlam1}
V^{(1)}(\z)\sim \Lambda_{j}^{(1)} \equiv -4\pi T^{(0)}\sum_{i=1}^NG_{ji}\calC_i+\chi^{(1)},\quad |\z|\rightarrow \infty .
\end{equation}
Proceeding along analogous lines to the case $n=0$, we find that
\begin{equation}
V^{(1)}(\z)=\left (1-\frac{\calC_j}{\z}\right )\Lambda_{j}^{(1)}
\end{equation}
with $\calC_j$ defined in equation (\ref{Fj}). It follows that the outer solution $T^{(2)}$ has the singular behavior
\begin{equation}
T^{(2)}\sim -\frac{\calC_j\Lambda_{j}^{(1)}}{|\x-\x_j|} \mbox{ as } \x\rightarrow \x_j.
\end{equation}
That is, $T^{(2)} $ satisfies the inhomogeneous equation
\begin{align}
D\nabla^2 T^{(2)}(\x) &=4\pi \sum_{j=1}^N\calC_j\Lambda_{j}^{(1)}\delta(\x-\x_j), \ \x \in\Omega; \quad \nabla T^{(2)} (\x)\cdot \n=0,\quad \x \in \partial \Omega.
\end{align}
Integrating this equation with respect to $\x\in \Omega$ and using the divergence theorem implies that
\begin{equation}
\sum_{j=1}^N\calC_j \left (-4\pi T^{(0)}\sum_{i=1}^NG_{ji}\calC_i+\chi^{(1)}\right )=0,
\end{equation}
where we have used equation (\ref{Tlam1}). That is,
\begin{align}
\chi^{(1)}=\frac{4\pi T^{(0)}}{N\overline{\calC}} \sum_{i,j=1}^N \calC_jG_{ji}\calC_i.
\end{align}

In summary, the outer solution for the unconditional MFPT is given by
\begin{align}
\label{Tf}
T(\x_0)&\sim \frac{|\Omega|}{4\pi D\epsilon N\overline{\calC}}\bigg \{1+ 4\pi \epsilon \sum_{i,j=1}^N\frac{1}{N\overline{\calC} } \calC_jG_{ji}\calC_i -4\pi \epsilon \sum_{j=1}^N\calC_jG(\x_0,\x_j) \bigg \}+O(\epsilon).
\end{align}
Again we recover previous results in the dual limits $\gamma \rightarrow \infty$ and $\kappa_j\rightarrow \infty$, $j=1,\ldots,N$, since $\calC_j\rightarrow \rho_j$ \cite{Coombs15,Bressloff21B}. As in the case of the splitting probabilities, $\calC_j$ acts as an effective capacitance. Finally note that for a single target, equation (\ref{Tf}) reduces to
\begin{align}
\label{Tf1}
T(\x_0)&\sim \frac{|\Omega|}{4\pi D\epsilon  {\calC}_1}+\frac{|\Omega|}{D}[R(\x_1,\x_1)-G(\x_0,\x_1) +O(\epsilon).
\end{align}
and the $O(1)$ contribution becomes independent of the parameters $(D_1,\kappa_1,\alpha_1,\rho_1)$.

\section{Example: triplet of targets in the unit sphere}

\begin{figure}[t!]
\centering
\includegraphics[width=11cm]{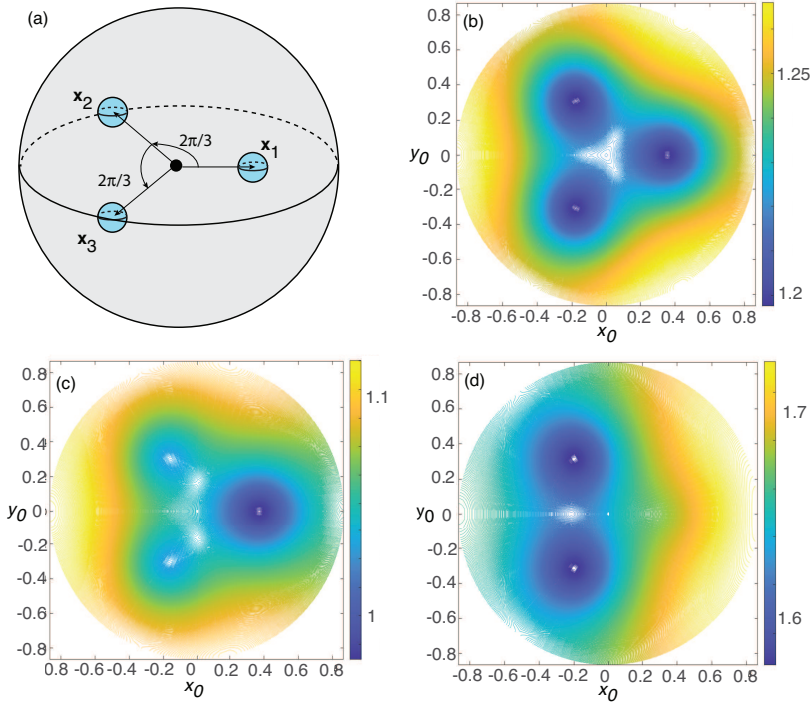} 
\caption{MFPT for narrow capture in the unit sphere. (a) A triplet of spherical traps are evenly distributed in the horizontal mid-plane at a radial distance $a= 0.5$ from the center. (b) Contour plot of MFPT $T(\x)$ given by equation (\ref{Tf}) as a function of the initial position $\x_0=(x_0,y_0,\cos \theta_0)$ with $\theta_0=\pi/3$ fixed. The three traps are identical with $D_j=\rho_j=1$, $\alpha_j=0.5$ and $\kappa_j=10$ for $j=1,2,3$. (c) Same as (b) except that $\alpha_1 =0.1$. (d) Same as (b) except that $\alpha_1=0.9$. Other parameters are $D=1$, $\gamma =1$ and $\epsilon =0.05$.}
\label{fig4}
\end{figure}

\begin{figure}[t!]
\centering
\includegraphics[width=12cm]{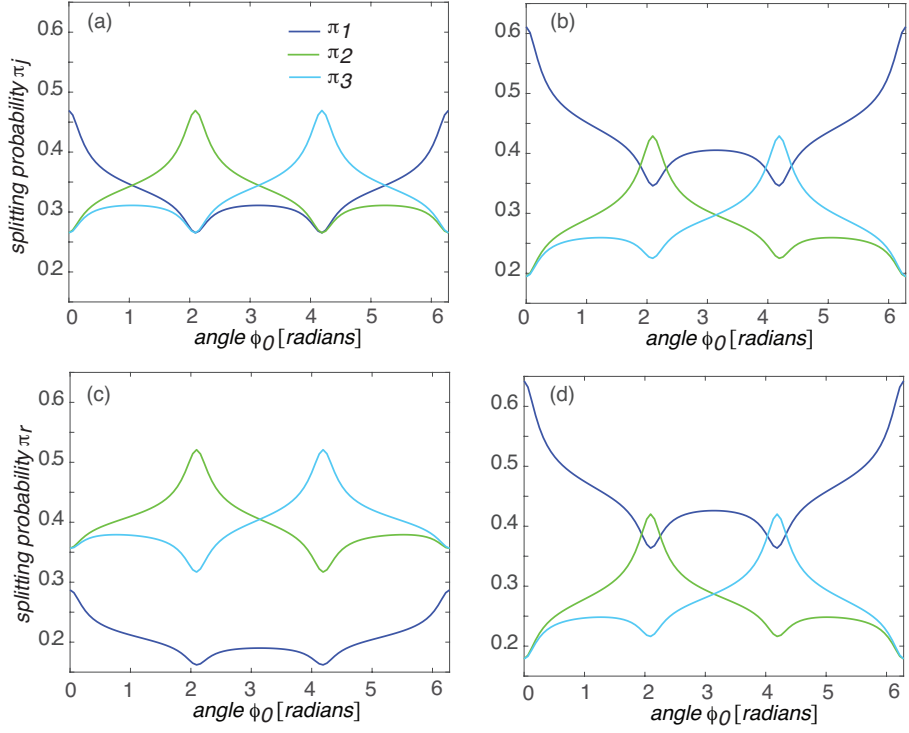} 
\caption{Splitting probabilities $\pi_j(\x_0)$, $j=1,2,3$, for narrow capture in the unit sphere with the triplet of traps shown in Fig. \ref{fig4}(a).  The initial position is taken to be in the $(x_0,y_0)$-plane, $\x_0=(\rho_0\ \cos \phi_0,\rho_0\sin\phi_0,0)$ with $\rho_0=0.6$. (a) Identical traps with the same parameter values as Fig. \ref{fig4}(b). (b) Same as (a) except that $\alpha_1=0.1$. (c) Same as (a) except that $\kappa_1=0.3$. (d) Same as (a) except that $\kappa_1=100$. Other parameters are $D=1$, $\gamma =1$ and $\epsilon =0.05$}
\label{fig5}
\end{figure}

\begin{figure}[t!]
\centering
\includegraphics[width=12cm]{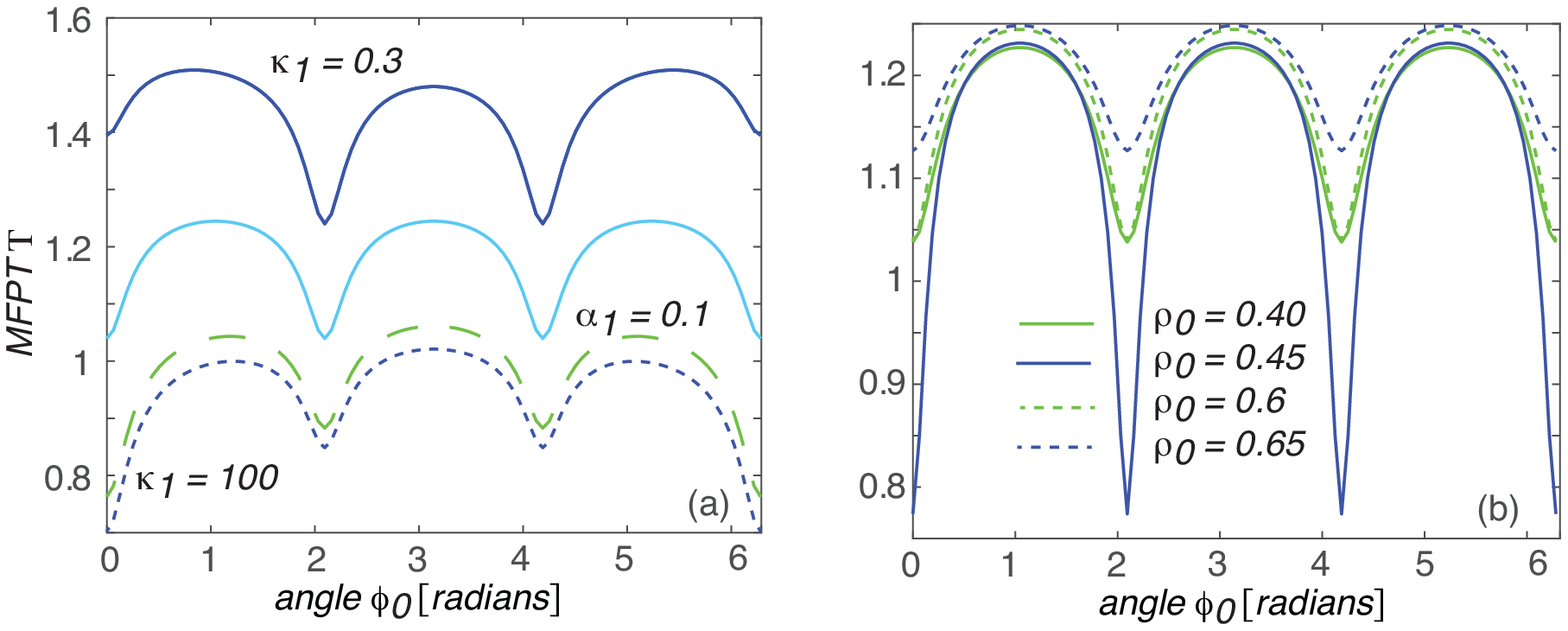} 
\caption{MFPT $T(\x_0)$ as a function of $\phi_0$ with $\x_0=(\rho_0\ \cos \phi_0,\rho_0\sin\phi_0,0)$. (a) Effect of changing one of the parameters of the first trap. Baseline levels are as in Fig. \ref{fig4}. (b) Identical traps as in Fig. \ref{fig4}(b) and various $\rho_0$. Other parameters are $D=1$, $\gamma =1$ and $\epsilon =0.05$}
\label{fig6}
\end{figure}

For the sake of illustration, suppose that the search domain $\Omega$ is the unit sphere and that it contains a triplet of targets evenly distributed in the horizontal mid-plane, see Fig. \ref{fig4}(a).
The 3D Neumann Green's function in the unit sphere is known explicitly \cite{Cheviakov11}:
\begin{align}
G_0(\x,\hxi)&=\frac{1}{4\pi |\x-\hxi|}+\frac{1}{4\pi |\x|r'} +\frac{1}{4\pi }\ln \left (\frac{2}{1-|\x||\hxi|\cos \theta+|\x|r'}\right ) 
\nonumber \\
&\quad +\frac{1}{6|\Omega|}(|\x|^2+|\hxi|^2)-\frac{7}{10\pi},
\end{align}
where $|\Omega|=4\pi/3$, and
\[\cos \theta =\frac{\x\cdot \hxi}{|\x||\hxi|},\quad \x'=\frac{\x}{|\x|^2},\quad r'=|\x'-\hxi|.
\]
The final constant is chosen so that $\int_{\Omega}G(\x,\hxi)d\x=0$. We assume that all traps have the same radius $\epsilon \overline{\rho}$ and take the positions of the traps to be
$\x_1=(a,0,0)$, $\x_2 =a(\cos 2\pi/3,\sin 2\pi/3,0)$ and $\x_2 =a(\cos 4\pi/3,\sin 4\pi/3,0)$ with $2\epsilon \overline{\rho} <a <1$.
In Fig. \ref{fig4}(b-d) we show contour plots of the MFPT $T(\x_0)$ given by equation (\ref{Tf}) for $N=3$. We display $T(\x_0)$ as a function of the $x_0$ and $y_0$ components of the initial position $\x_0$ for fixed $z_0$, namely, $\x_0=(x_0,y_0,\cos\pi/3)$. The traps are taken to be identical except for different choices of the bias $\alpha_1$ in the first trap. It can be seen that if $\alpha_1 < \alpha_2=\alpha_3$ then the MFPT is smaller if it starts in a neighborhood of the first trap. On the other hand, if $\alpha_1>\alpha_2=\alpha_3$ then the MFPT is reduced if it starts in a neighborhood of the second or third trap. Finally, note that  the $\x_0$ dependence of $T(\x_0)$ is particularly weak, since we are considering $\x_0$ away from the horizontal mid-plane.

Suppose that the initial condition is now taken to be in the $(x_0,y_0)$-plane such that $\x_0=(\rho_0  \cos \phi,\rho_0 \sin\phi_0,0)$ with $\rho_0$ fixed. In Fig. \ref{fig5} we plot the splitting probabilities $\pi_j$, $j=1,2,3$, as a function of the angular coordinate $\phi_0 \in [0,2\pi]$ for different parameter values of the first trap. Clearly $\pi_j$ has a maximum when $\phi_0=\phi_j$ and minima at $\phi_0=\phi_k$, $k\neq j$, where $\phi_k$ is the angular coordinate of the $k$th trap. For identical traps, the splitting probabilities are identical up to a $2\pi/3$ shift in $\phi_0$. However, as expected, decreasing $\alpha_1$ or increasing $\kappa_1$ increases $\pi_1$ and reduces $\pi_2$ and $\pi_3$. Analogous plots for the MFPT are presented in Fig. \ref{fig6}.

\section{Alternative parameter scalings}

So far we have assumed the scalings (\ref{scale1}) with $\gamma',\kappa_j'=O(1)$. In this section we perform an asymptotic analysis of the splitting probabilities in the slow absorption regime $\gamma=O(1/\epsilon)$ and the low permeability regime $\kappa_j=O(1)$, respectively\footnote{Under the scaling $\gamma=O(1/\epsilon)$, the effective absorption rate in stretched coordinates is $\epsilon^2 \gamma=O(\epsilon)$. Similarly, if $\kappa=O(1)$, then the effective permeability in stretched coordinates is $\epsilon \kappa_j=O(\epsilon)$.}. In both cases we find that the $O(1)$ terms in the asymptotic expansions agree with equation (\ref{bigp}) under the approximations (\ref{app1}) and (\ref{app2}), respectively.  However, the $O(\epsilon)$ terms are now constant so the $\x_0$-dependence of $\pi_k(\x_0)$ first occurs at $O(\epsilon^2)$. Hence, the narrow capture problem can no longer be reduced to the totally absorbing case in an appropriate limit. (Analogous results hold for the MFPT.)

\subsection{Slow absorption} Suppose that 
\begin{equation}
  \label{scale2}
  \gamma=\frac{\gamma'}{\epsilon},\quad  \kappa_j= \frac{\kappa'_j}{\epsilon} .
  \end{equation} 
  The inner solutions are now expanded as
\begin{equation}
 U_{k}=\pi_k^{(0)}+\epsilon U_{k}^{(1)}+O(\epsilon^2),\quad V_k=\pi_{k}^{(0)}+\epsilon V_{k}^{(1)}+O(\epsilon^4),
 \end{equation}
 with
 \begin{subequations} 
\label{0innersplit}
\begin{align}
 &D\nabla^2  V_k^{(n)}(\z)=0,\ |\z|>\rho_j,\\
  &D_j\nabla^2 U_k^{(n)}(\z) +\gamma' (\delta_{j,k}\delta_{n,1}- U_k^{(n-1)}(\z))=0,\ |\z|<\rho_j,\ n\geq 2,\\
\alpha_jD\nabla V_k^{(n)}(\z) \cdot \n_j  &= (1-\alpha_j) D_j\nabla U_k^{(n)}(\z) \cdot \n_j,\nonumber \\
& = \kappa_j' \alpha_j(1-\alpha_j) [U_k^{(n)}(\z)-V_k^{(n)}(\z)] ,\ |\z|=\rho_j.
\end{align}
\end{subequations}
Following along similar lines to section 3.1, we find that
\begin{equation}
\label{0sola}
U_k^{(1)}(\rho)=B_{jk}^{(1)} -\frac{\gamma' \rho^2}{6D_j}[\delta_{j,k}-\pi_k^{(0)}],\quad V_k^{(1)}(\rho)=\frac{A_{jk}^{(1)}}{\rho} +\pi_k^{(1)}.
\end{equation}
Substituting into (\ref{0innersplit}c) yields
\begin{subequations}
\begin{align}
-\frac{\alpha_jDA_{jk}^{(1)}}{\rho_j}&=-\frac{\gamma'(1-\alpha_j)\rho_j^2}{3} [\delta_{j,k}-\pi_k^{(0)}],\\
-\frac{\alpha_jDA_{jk}^{(1)}}{\rho_j}&=\kappa_j'\alpha_j(1-\alpha_j)\left [A_{jk}^{(1)}+\rho_j\pi_k^{(1)}+\frac{\gamma' \rho_j^3}{6D_j}[\delta_{j,k}-\pi_k^{(0)}]- \rho_j B_{jk}^{(1)} \right ].
\end{align}
\end{subequations}
Hence,
\begin{equation}
\label{Ajk}
A_{jk}^{(1)}=\calC_j^{\gamma}[\delta_{j,k}-\pi_k^{(0)}],\quad \calC_j^{\gamma}=\frac{\gamma'(1-\alpha_j)|\calU_j^0|}{4\pi D\alpha_j},
\end{equation}
and
\begin{align}
\label{Bj}
\rho_j  B_{jk}^{(1)}&=\left (1+\frac{D}{\kappa'_j \rho_j (1-\alpha_j)}   +\frac{\alpha_j D}{2(1-\alpha_j)D_j}   \right )A_{jk}^{(1)}+\rho_j \pi_k^{(1)}.
\end{align}

The outer solution has the asymptotic expansion $\pi_k(\x)\sim \pi_k^{(0)} +\epsilon \pi_k^{(1)}+\epsilon^2 \pi_k^{(2)}(\x)+O(\epsilon^3)$ with $\pi_k^{(0,1)}$ constants, and
\[\pi_k^{(2)} (\x)\sim \frac{A_{jk}^{(1)}}{|\x-\x_j|}\quad \mbox{as } \x\rightarrow \x_j.\]
Proceeding in a similar fashion to section 3.2, we have
\begin{equation}
\label{pi3}
\pi_k^{(2)}(\x)=4\pi \sum_{j=1}^NA_{jk}^{(1)}G(\x,\x_j)+\chi_{k}^{(2)},\quad \sum_{j=1}^NA_{jk}^{(1)}=0.
\end{equation}
In particular, $\pi_k^{(0)}$ has the form (\ref{pi0}) under the mapping $\calC_j\rightarrow \calC_j^{\gamma}$:
\begin{equation}
\label{gampi0}
\pi_k^{(0)} =\frac{\calC^{\gamma}_k}{N\overline{\calC}^{\gamma}} ,\quad \overline{\calC}^{\gamma}=\frac{1}{N}\sum_{j=1}^N \calC_j^{\gamma}.
\end{equation} 
Since $\sum_{k=1}^N\pi_k^{(0)}=1$ we see from equation (\ref{Ajk}) that $\sum_{k=1}^NA_{jk}^{(1)}=0$.

Finally, $\pi_k^{(1)}$ is determined by considering the next term in the inner solution along the lines of section 3.3. First, note that
\begin{equation}
V_k^{(2)}(\z)=\frac{A_{jk}^{(2)}}{\rho}+\Lambda_{jk}^{(2)} ,\quad \Lambda_{jk}^{(2)} \equiv 4\pi \sum_{i=1}^NG_{ji}A_{ik}^{(1)}+\chi_{k}^{(2)}.
\end{equation}
On the other hand,
\begin{equation}
 D_j\nabla^2 U_k^{(2)}(\z) -\gamma' U_k^{(1)}(\z)=0,\ |\z|<\rho_j
 \end{equation}
 Setting $U_k^{(2)}(\z) =\phi_k^{(2)}(\rho)/\rho$, we have
\begin{align}
 D_j\frac{\partial^2\phi_k^{(2)}(\rho)}{\partial \rho^2} -\gamma' \rho B_{jk}^{(1)} +\frac{{\gamma'}^2 \rho^3}{6D_j}[\delta_{j,k}-\pi_k^{(0)}]&=0 ,\  \rho <\rho_j,
\end{align}
We thus have
\begin{equation}
U_k^{(2)}(\rho)= -\frac{{\gamma'}^2 \rho^4}{5! D_j^2}[\delta_{j,k}-\pi_k^{(0)}]+\frac{\gamma' \rho^2 B_{jk}^{(1)}}{6D_j}+B_{jk}^{(2)}.
\end{equation}
Substituting the resulting solutions for $ U_k^{(2)}$ and $ V_k^{(2)}$ into (\ref{0innersplit}c) yields
\begin{subequations}
\begin{align}
-\frac{\alpha_jDA_{jk}^{(2)}}{\rho_j}&=(1-\alpha_j)D_j\left (-\frac{{\gamma'}^2 \rho_j^5}{30 D_j^2}[\delta_{j,k}-\pi_k^{(0)}]+\frac{\gamma' \rho_j^3 B_{jk}^{(1)}}{3D_j}\right )\\
-\frac{\alpha_jDA_{jk}^{(2)}}{\rho_j}&=\kappa_j'\alpha_j(1-\alpha_j)\bigg [A_{jk}^{(2)}+\rho_j \Lambda_{jk}^{(2)}+\frac{{\gamma'}^2 \rho_j^5}{5! D_j^2}[\delta_{j,k}-\pi_k^{(0)}]-\frac{\gamma' \rho_j^3 B_{jk}^{(1)}}{6D_j}- \rho_j B_{jk}^{(2)} \bigg ].
\end{align}
\end{subequations}
The inner solution $V_k^{(2)}$ determines the singular behavior of the outer solution $\pi_k^{(3)}$ and the resulting diffusion equation implies that $\sum_{j=1}^N A_{jk}^{(2)}=0$. Hence, 
\begin{align}
0&=\sum_{j=1}^N\frac{(1-\alpha_j)D_j}{\alpha_j D}\left (-\frac{{\gamma'}^2 \rho_j^6}{30 D_j^2}[\delta_{j,k}-\pi_k^{(0)}]+\frac{\gamma' \rho_j^4 B_{jk}^{(1)}}{6D_j}\right )\nonumber \\
&=\sum_{j=1}^N\calC_j^{\gamma}\left (\rho_jB_{jk}^{(1)} -\frac{\alpha_j D}{5(1-\alpha_j)D_j}A_{jk}^{(1)}\right ).
\end{align}
Substituting for $B_{jk}^{(1)}$ using equation (\ref{Bj}) yields
\begin{align}
 \pi_k^{(1)}&=-\frac{1}{\Theta^{\gamma}}\sum_{j=1}^N\calC_j^{\gamma}\left (1+\frac{D}{\kappa'_j \rho_j (1-\alpha_j)}   +\frac{3\alpha_j D}{10(1-\alpha_j)D_j}   \right )\calC_j^{\gamma}[\delta_{j,k}-\pi_k^{(0)}],
\end{align}
where $\Theta^{\gamma}=\sum_{j=1}^N\rho_j \calC_j^{\gamma}$. 
Note that $\sum_{k=1}^N\pi_k^{(1)}=0$.

\subsection{Low permeability}
Now suppose that $\gamma=\gamma'/\epsilon^2$, whereas $\kappa_j=O(1)$.
The inner solution is
 \begin{subequations} 
\label{1innersplit}
\begin{align}
&D\nabla^2  V_k^{(n)}(\z)=0,\ |\z|>\rho_j,\\
& D_j\nabla^2 U_k^{(n)}(\z) + \gamma' (\delta_{j,k}\delta_{n,0}-U_k^{(n)}(\z))=0,\ |\z|<\rho_j,\\
\alpha_jD\nabla V_k^{(n)}(\z) \cdot \n_j  &= (1-\alpha_j) D_j\nabla U_k^{(n)}(\z) \cdot \n_j\nonumber \\
& = \kappa_j \alpha_j(1-\alpha_j) [U_k^{(n-1)}(\z)-V_k^{(n-1)}(\z)] ,\ |\z|=\rho_j,
\end{align}
\end{subequations}
with $U_k^{(-1)},V_k^{(-1)}\equiv 0$. Equations (\ref{1Dsola}) still hold for $U_k^{(0)},V_k^{(0)}$ but the boundary conditions now take the form
\[\nabla V_k^{(0)}(\z) \cdot \n_j  = \nabla U_k^{(0)}(\z) =0.\]
It follows that $U_k^{(0)}= \delta_{j,k}$ and $V_k^{(0)}=\pi_k^{(0)}$. In addition, as in the previous example, the outer solution has the expansion
$\pi_k(\x)\sim \pi_k^{(0)} +\epsilon \pi_k^{(1)}+\epsilon^2 \pi_k^{(2)}(\x)+O(\epsilon^3)$ with $\pi_k^{(0,1)}$ constants.

The $O(\epsilon)$ inner solution has the same form as equation (\ref{11Dsola}): 
\begin{equation}
U_k^{(1)}(\rho)= B_{jk}^{(1)}\frac{\sinh(\sqrt{\gamma'/D_j}\rho)}{\rho} ,\quad V_k^{(1)}(\rho)=\frac{A_{jk}^{(1)}}{\rho}+\pi_k^{(1)} .
\end{equation}
The corresponding boundary conditions are 
\begin{align}
\alpha_jD\nabla V_k^{(1)}(\z) \cdot \n_j  &= (1-\alpha_j) D_j\nabla U_k^{(1)}(\z) \cdot \n_j\nonumber \\
& = \kappa_j \alpha_j(1-\alpha_j) [\delta_{j,k}-\pi_k^{(0)}] ,\ |\z|=\rho_j,
\end{align}
This implies that
\begin{subequations}
\begin{align}
-\frac{\alpha_jDA_{jk}^{(1)}}{\rho_j}&=(1-\alpha_j)B_{jk}^{(1)}\left \{\sqrt{\gamma' D_j}\cosh(\sqrt{\gamma'/D_j}\rho_j) - \frac{D_j}{\rho_j}\sinh(\sqrt{\gamma'/D_j}\rho_j) \right \},\\
-\frac{\alpha_jDA_{jk}^{(1)}}{\rho_j^2}&= \kappa_j \alpha_j(1-\alpha_j) [\delta_{j,k}-\pi_k^{(0)}].
\end{align}
\end{subequations}
Hence,
\begin{equation}
A_{jk}^{(1)}=-\calC_j^{\kappa}[\delta_{j,k}-\pi_k^{(0)}],\quad \calC_j^{\kappa}=\frac{\kappa_k (1-\alpha_j)|\partial \calU_j^0|}{4\pi D},\quad B_{jk}^{(1)}=- \calG_j A_{jk}^{(1)},
\end{equation}
where
\begin{equation}
 \calG_j =\frac{\alpha_jD}{(1-\alpha_j)\rho_j} \left \{\sqrt{\gamma' D_j}\cosh(\sqrt{\gamma'/D_j}\rho_j) - \frac{D_j}{\rho_j}\sinh(\sqrt{\gamma'/D_j}\rho_j) \right \}^{-1}
\end{equation}
The outer solution $\pi_k^{(2)}(\x)$ is then given by equation (\ref{pi3}) so that 
\begin{equation}
\label{kappi0}
\pi_k^{(0)} =\frac{\calC^{\kappa}_k}{N\overline{\calC}^{\kappa}} ,\quad \overline{\calC}^{\kappa}=\frac{1}{N}\sum_{j=1}^N \calC_j^{\kappa}.
\end{equation} 

The $O(\epsilon^2)$ inner solution has the form
\begin{equation}
U_k^{(2)}(\rho)= B_{jk}^{(2)}\frac{\sinh(\sqrt{\gamma'/D_j}\rho)}{\rho} ,\quad V_k^{(2)}(\rho)=\frac{A_{jk}^{(2)}}{\rho}+\Lambda_{jk}^{(2)} ,
\end{equation}
with
\begin{equation}  \Lambda_{jk}^{(2)} \equiv 4\pi \sum_{i=1}^NG_{ji}A_{ik}^{(1)}+\chi_{k}^{(2)}.
\end{equation}
The corresponding boundary conditions are 
\begin{align}
\alpha_jD\nabla V_k^{(2)}(\z) \cdot \n_j  &= (1-\alpha_j) D_j\nabla U_k^{(2)}(\z) \cdot \n_j\nonumber \\
& = \kappa_j \alpha_j(1-\alpha_j) [\delta_{j,k}-\pi_k^{(0)}] ,\ |\z|=\rho_j,
\end{align}
This implies that
\begin{subequations}
\begin{align}
-\frac{\alpha_jDA_{jk}^{(2)}}{\rho_j}&=(1-\alpha_j)B_{jk}^{(2)}\left \{\sqrt{\gamma' D_j}\cosh(\sqrt{\gamma'/D_j}\rho_j) - \frac{D_j}{\rho_j}\sinh(\sqrt{\gamma'/D_j}\rho_j) \right \},\\
-\frac{\alpha_jDA_{jk}^{(2)}}{\rho_j^2}&= \kappa_j\alpha_j(1-\alpha_j)\left [A_{jk}^{(1)}+\pi_k^{(1)}\rho_j  - B_{jk}^{(1)}\sinh(\sqrt{\gamma'/D_j}\rho_j) \right ].\end{align}
\end{subequations}
Finally, the diffusion equation for $\pi_k^{(3)}$ implies that $\sum_{j=1}^NA_{jk}^{(2)}=0$ and, thus,
\begin{align}
0&=\sum_{j=1}^N\calC_j^{\kappa}\left (A_{jk}^{(1)}+\pi_k^{(1)}\rho_j  - B_{jk}^{(1)}\sinh(\sqrt{\gamma'/D_j}\rho_j) \right ).
\end{align}
Rearranging shows that
\begin{align}
 \pi_k^{(1)}=\frac{1}{\Theta^{\kappa}}\sum_{j=1}^N\calC_j^{\kappa} \left (1 + \calG_j \sinh(\sqrt{\gamma'/D_j}\rho_j) \right )\calC_j^{\kappa} [\delta_{j,k}-\pi_k^{(0)}],
\end{align}
with $ {\Theta}^{\kappa}=\sum_{j=1}^N\calC_j^{\kappa}\rho_j$.

\setcounter{equation}{0}
\section{Partially reactive surfaces} 
Taking $\alpha_j=0$ leads to the narrow capture problem of equations (\ref{masterR}), with $\kappa_j$ the effective constant reaction rate for the Robin boundary condition at the $j$th trap, see Fig. \ref{fig2}(c). For simplicity we will assume that $\kappa_j=\kappa_0$ for all $j=1,\ldots,N$. We can then  reformulate the Robin boundary conditions using a probabilistic interpretation based on a boundary local time \cite{Grebenkov20a,Grebenkov22,Bressloff22b,Bressloff22c}. The latter is a Brownian functional that keeps track of the amount of time a particle spends in a local neighborhood of the total interior boundary $\partial \calU_a$ with $\calU_a=\sum_{j=1}^N\calU_j$. Let $\X_t \in\Omega\backslash \calU_a$ represent the position of the diffusing particle at time $t$. If the surface $\partial \calU_a$ is totally reflecting then we can define the boundary local time according to
\begin{equation}
\label{loc}
\ell_t=\lim_{h\rightarrow 0} \frac{D}{h} \int_0^t\Theta(h-\mbox{dist}(\X_{\tau},\partial \calU_a))d\tau,
\end{equation}
where $\Theta$ is the Heaviside function. Note that $\ell_t$ has units of length due to the additional factor of $D$. Let $P(\x,\ell,t|\x_0)$ denote the joint probability density or propagator for the pair $(\X_t,\ell_t)$ and introduce the stopping time 
\begin{equation}
\label{Tell0}
{\mathcal T}=\inf\{t>0:\ \ell_t >\widehat{\ell}\},
\end{equation}
 with $\widehat{\ell}$ an exponentially distributed random variable that represents a stopping local time. That is, $\P[\widehat{\ell}>\ell]=\e^{-z_0\ell}$
with $z_0=\kappa_0/D$. (Roughly speaking, the stopping time ${\mathcal T}$ is a random variable that specifies the time of absorption, which is determined by the instant at which the local time $\ell_t$ crosses the random threshold  $\widehat{\ell}$.) The marginal density $p(\x,t|\x_0)$ satisfying equations (\ref{masterR}) can be related to the propagator  $P(\x,\ell,t|\x_0)$ as follows: 
\begin{align*}
p(\x,t|\x_0)d\x&=\P[\X_t \in (\x,\x+d\x), \ t < {\mathcal T}|\X_0=\x_0].\\
&=\P[\X_t \in (\x,\x+d\x), \ \ell_t < \widehat{\ell}|\X_0=\x_0]\\
&=\int_0^{\infty} d\ell \ z_0\e^{-z_0\ell}\P[\X_t \in (\x,\x+d\x), \ \ell_t < \ell |\X_0=\x_0]\\
&=\int_0^{\infty} d\ell \ z_0 \e^{-z_0\ell}\int_0^{\ell} d\ell' [P(\x,\ell',t|\x_0)d\x].
\end{align*}
We have used the fact that $\ell_t$ is a nondecreasing process so that the condition $t < {\mathcal T}$ is equivalent to the condition $\ell_t <\widehat{\ell}$. Using the identity
\begin{equation}
\int_0^{\infty}d\ell \ f(\ell)\int_0^{\ell} d\ell' \ g(\ell')=\int_0^{\infty}d\ell' \ g(\ell')\int_{\ell'}^{\infty} d\ell \ f(\ell)
\label{fg}
\end{equation}
for arbitrary integrable functions $f,g$, it follows that
\begin{equation}
\label{bob}
p(\x,t|\x_0)=\int_0^{\infty} \e^{-z_0\ell}P(\x,\ell,t|\x_0)d\ell\equiv \widetilde{P}(\x,\gamma,t|\x_0).
\end{equation}

Hence, the marginal density $p(\x,t|\x_0)$ is equivalent to the Laplace transform $\widetilde{P}(\x,z_0,t|\x_0)$ of the propagator with respect to the local time $\ell$. In other words, $\widetilde{P}(\x,z,t|\x_0)$ satisfies the Robin BVP (\ref{masterR}) with $\kappa_0=z_0D$.
The advantage of this formulation is that one can consider a more general probability distribution $\Psi(\ell) = \P[\widehat{\ell}>\ell]$ for the stopping local time $\widehat{\ell}$ such that \cite{Grebenkov20a,Grebenkov22,Bressloff22b,Bressloff22c}.
  \begin{equation}
  \label{Boo}
  p^{\Psi}(\x,t|\x_0)=\int_0^{\infty} \Psi(\ell)P(\x,\ell,t|\x_0)d\rho=\int_0^{\infty} \Psi(\ell){\mathcal L}^{-1}\widetilde{P}(\x,z,t|\x_0)d\rho,
  \end{equation}
  where ${\mathcal L}^{-1}$ denotes the inverse Laplace transform with respect to $z$.
The steps of the general encounter-based method for a Brownian particle can be summarized as follows: 
\medskip

\noindent 1. Solve the Robin BVP for the probability density $p(\x,t|\x_0)$ in the case of a constant rate of absorption $\kappa_0$.
\medskip

\noindent 2. Identify $p(\x,t|\x_0)$ with the Laplace transform $\widetilde{P}(\x,z,t|\x_0)$ of the local time propagator, with $\kappa_0=zD$.
\medskip

\noindent 3. Invert the Laplace transform with respect to $z$ to obtain $P(\x,\ell,t|\x_0)$.
\medskip

\noindent 4. Calculate the general marginal density according to (\ref{Boo}).
\medskip

\noindent An alternative approach would be to solve the BVP for the local time propagator directly, which is how we proceeded in Ref. \cite{Bressloff22a} for the narrow capture problem in $\R^3$. Here we will work directly with the asymptotic expansions of the splitting probabilities and MFPT in the bounded domain $\Omega$. 

First, consider the case of constant reactivity $\kappa_0$. The total probability flux into the partially absorbing boundary $\calU_k$ is then
\begin{equation}
J_k(\x_0,t) =D\int_{\calU_k} \nabla p(\x,t|\x_0)\cdot \n_j d\x,\quad \x_0\in \Omega\backslash \calU_a,
\end{equation}
with $p$ the solution to the BVP (\ref{masterR}). 
Hence, the corresponding splitting probabilities and unconditional MFPT are $\pi_k(\x_0)=\widetilde{J}_k(\x_0,0)$ and 
$ T(\x_0)=-\partial_s \widetilde{J}_k(\x_0,0)$.
Suppose that we include $z=\kappa_0/D$ as an additional variable in the flux by writing $\widetilde{J}_k=\widetilde{J}_k(\x_0,z,s)$, and set
 \begin{equation}
\pi_k(\x_0,z)=\widetilde{J}_k(\x_0,z,0),\quad T(\x_0,z)=-\partial_s \widetilde{J}_k(\x_0,z,0).
\end{equation}
Using the above encounter-based method, the splitting probabilities and MFPT for a general reaction scheme can then be written as
 \begin{equation}
 \label{piPsi}
\pi_k^{\Psi}(\x_0)=\int_0^{\infty} \Psi(\ell) {\mathcal L}^{-1}\pi_k(\x_0,z)d\ell,\quad T^{\Psi}(\x_0)=-\int_0^{\infty} \Psi(\ell) {\mathcal L}^{-1}T(\x_0,z)d\ell.
\end{equation}
It remains to determine the explicit dependence of $\pi_k$ and $T$ on $z$. We now make the crucial observation that the flux $\widetilde{J}_k$ is related to the flux $\widetilde{\calJ}_k$ according to equations (\ref{piRobin}) and (\ref{TRobin}). That is, although the interior of each trap no longer plays any role, we can still use the solutions of the BVPs (\ref{BVPsplit}) and (\ref{BVPT}) to determine $\pi_k(\x_0,z)$ and $T(\x_0,z)$ for partially absorbing surfaces with constant reactivity $\kappa_0=zD$ by setting $\alpha_j=0$ in equations (\ref{bigp}) and (\ref{Tf}). This is equivalent to setting
\begin{equation}
\calC_j={\rho_j}\left [\frac{D}{\kappa_0\rho_j}  +1\right ]^{-1}=\frac{z\rho_j}{\rho_j^{-1}+z}=\rho_j-\frac{1}{\rho_j^{-1}+z}\equiv \calC_j(z).
\end{equation}
It follows that the construction of the asymptotic solutions for general $\Psi$ reduces to the problem of finding the inverse Laplace transforms of the terms involving the capacitances $\calC_k(z)$. (We are using the fact that an asymptotic expansion commutes with integral operators.)

As a simple example, suppose that $\rho_j=\overline{\rho}$ for all $j=1,\ldots,N$. Equation (\ref{bigp}) reduces to the simpler form
\begin{align}
\label{piz}
\pi_k(\x_0,z)&\sim \frac{1}{N} \left \{1+ 4\pi \epsilon  {\calC}(z)\left (\sum_{i,j=1}^N\frac{1}{N} G_{ji} + \sum_{j=1}^N [G(\x_0,\x_k)-G(\x_0,\x_j)-G_{jk}]\right )\right \}\nonumber \\
&\hspace{3cm}+O(\epsilon^2).
\end{align}
with
\begin{equation}
\label{calz}
\calC(z)=\overline{\rho}-\frac{1}{\overline{\rho}+z},\quad C(\ell)\equiv [{\mathcal L}^{-1} \calC](\ell)=\overline{\rho}\delta(\ell)-\e^{-\overline{\rho}\ell}.
\end{equation}
Substituting equation (\ref{piz}) into the first equation of (\ref{piPsi}) and using (\ref{calz}) yields the generalized asymptotic expansion
\begin{align}
\pi_k^{\Psi}(\x_0)&\sim \frac{1}{N}\int_0^{\infty}\Psi(\ell)  \bigg\{\delta(\ell)+ 4\pi \epsilon \bigg[\overline{\rho}\delta(\ell)-\e^{-\ell/\overline{\rho} }\bigg ]\bigg (\sum_{i,j=1}^N\frac{1}{N} G_{ji} \nonumber \\
&\quad + \sum_{j=1}^N [G(\x_0,\x_k)-G(\x_0,\x_j)-G_{jk}]\bigg )\bigg\}+O(\epsilon^2)\nonumber \\
&\sim \frac{1}{N}\int_0^{\infty}   \bigg\{1+ 4\pi \epsilon \bigg[\overline{\rho}-\widetilde{\Psi}(1/\overline{\rho})\bigg ]\bigg (\sum_{i,j=1}^N\frac{1}{N} G_{ji} \nonumber \\
&\quad + \sum_{j=1}^N [G(\x_0,\x_k)-G(\x_0,\x_j)-G_{jk}]\bigg )\bigg\}+O(\epsilon^2).
\end{align}
Similarly, equation (\ref{Tf}) becomes
\begin{align}
T(\x_0,z)&\sim \frac{|\Omega|}{ D}\bigg \{ \frac{1}{4\pi \epsilon {\calC}(z)}+  \sum_{i,j=1}^N  G_{ji}\ - \sum_{j=1}^N G(\x_0,\x_j) \bigg \}+O(\epsilon).
\end{align}
with
\begin{equation}
\label{calz2}
\calF(z)\equiv \calC(z)^{-1}=\frac{\overline{\rho}^{-1}+z}{z\overline{\rho}},\quad F(\ell)\equiv [{\mathcal L}^{-1} \calF](\ell)=\frac{1}{\overline{\rho}}\delta(\ell)-\frac{1}{\overline{\rho}^2}.
\end{equation}
Hence
\begin{align}
T^{\Psi}(\x_0)&\sim \frac{|\Omega|}{ D}\bigg \{ \frac{1}{4\pi \epsilon \overline{\rho}}\left (1+\frac{\overline{\Psi}}{\overline{\rho}}\right )+  \sum_{i,j=1}^N  G_{ji}\ - \sum_{j=1}^N G(\x_0,\x_j) \bigg \}+O(\epsilon),
\end{align}
where $\overline{\Psi}=\int_0^{\infty}\Psi(\ell)d\ell$. 

\begin{figure}[t!]
\centering
\includegraphics[width=8cm]{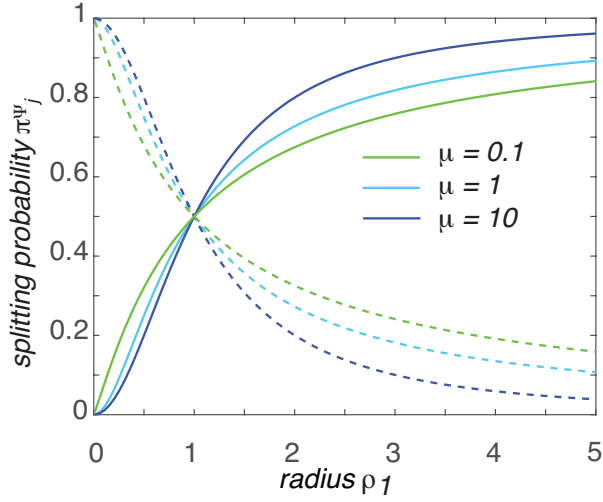} 
\caption{Pair of spherical traps with non-Markovian, partially absorbing surfaces $\partial \calU_j$ and radii $\rho_j$, $j=1,2$. Plots of the leading order contribution to the splitting probabilities $\pi_1^{\Psi}$ (solid curves) and $\pi_2^{\Psi}$ (dashed curves) as a function of $\rho_1$ with $\rho_2=1$. (The leading order terms are independent of $\x_0$.) The local time distribution $\psi(\ell)=-\Psi'(\ell)$ is taken to be the gamma distribution (\ref{psigam}) with parameters $(\mu,\eta)$ and $\eta=1$.}
\label{fig7}
\end{figure}

It is possible to extend the analysis to the case of inhomogeneous radii $\rho_i$, but the algebra is more involved, since one has to find the poles of potentially high-dimensional rational functions in order to calculate the inverse Laplace transforms. For example, suppose that we have two traps. The $O(1)$ contribution to $\pi_1$ is independent of $\x_0$ and takes the form
\begin{align}
\pi_1(z)&= \frac{\calC_1(z)}{\calC_1(z)+\calC_2(z)} =\frac{1}{1+\frac{\rho_2}{\rho_1}\frac{\rho_1^{-1}+z}{\rho_2^{-1}+z}}=\frac{\rho_1(\rho_2^{-1}+z)}{\rho_1^{-1}\rho_2+\rho_1\rho_2^{-1}+(\rho_1+\rho_2)z}\nonumber \\
&=\frac{\rho_1}{\rho_1+\rho_2}\frac{\rho_2^{-1}+z}{\Sigma_{12}+z},\end{align}
with
\begin{equation}
\Sigma_{12}\equiv \frac{\rho_1^{-1}\rho_2+\rho_1\rho_2^{-1}}{\rho_1+\rho_2}=\frac{1}{\rho_1}+\frac{1}{\rho_2}-\frac{2}{\rho_1+\rho_2}.
\end{equation}
Hence,
\begin{align}
{\mathcal L}^{-1}[\pi_1](\ell)=\frac{\rho_1}{\rho_1+\rho_2}\left [\delta(\ell) +[\rho_2^{-1}-\Sigma_{12}]\e^{-\Sigma_{12}\ell}\right ],
\end{align} 
so that
\begin{equation}
\pi_1^{\Psi}=\int_0^{\ell}\Psi(\ell) {\mathcal L}^{-1}[\pi_1](\ell)d\ell= \frac{\rho_1}{\rho_1+\rho_2}\left [1 +[\rho_2^{-1}-\Sigma_{12}]\widetilde{\Psi}(\Sigma_{12})\right ]
\end{equation}
Similarly, $\pi_2^{\Psi}$ is obtained under the interchange $\rho_1\leftrightarrow \rho_2$. It can be checked that $\pi_1^{\Psi} +\pi_1^{\Psi} =1$ since $\Sigma_{12}=\Sigma_{21}$. For the sake of illustration, suppose that $\psi(\ell)\equiv -\Psi'(\ell)$ is given by the gamma distribution:
\begin{equation}
\label{psigam}
\psi(\ell)=\frac{\eta(\eta\ell)^{\mu-1}\e^{-\eta \ell}}{\Gamma(\mu)},\  \mu >0,
\end{equation}
where $\Gamma(\mu)$ is the gamma function. The corresponding Laplace transforms are
\begin{equation}
\widetilde{\psi} (z)=\left (\frac{\eta}{\eta+z}\right )^{\mu},\quad \widetilde{\Psi}(z)=\frac{1-\widetilde{\psi}(z)}{z}.
\end{equation}
Note that if $\mu=1$ then $\psi$ reduces to the exponential distribution with constant reactivity $\kappa_0 = D\eta$. The parameter $\mu$ thus characterizes the deviation of $\psi(\ell)$ from the exponential case. If $\mu <1$ ($\mu>1$) then $\psi(\ell)$ decreases more rapidly (slowly) as a function of the local time $\ell$. The effective reaction time is given by the mean $\E[\ell]=\mu/\eta$. Example plots of the leading order contributions to $\pi_1^{\Psi}$ and $\pi_2^{\Psi}$ as a function of $\rho_1$ for $\rho_2=1$ are shown in Fig. \ref{fig7}. It can be seen that increasing $\mu$ reduces $\pi_1^{\Psi}$ when $\rho_1 <\rho_2$ and has the opposite effect when $\rho_1>\rho_2$.

\section{Discussion}

In this paper we developed a general theory of the 3D narrow capture problem for small spherical traps in a bounded domain, with the boundaries of the traps taken to be semipermeable. We used singular perturbation theory and Greens function methods to derive the leading order terms in the asymptotic expansions of the splitting probabilities and the unconditional MFPT. (Other quantities such as the conditional MFPTs and higher FPT  moments could have been analyzed in a similar fashion, see  Ref. \cite{Coombs15,Bressloff22}.) We also showed how previously studied versions of the narrow capture problem can be recovered in appropriate limits, including partially absorbing interior substrates \cite{Bressloff22}, totally absorbing surfaces \cite{Coombs15,Bressloff21B} and partially reactive surfaces \cite{Bressloff22a}. This established that the main difference between the various cases is the form of the effective capacitances $\calC_i$; the general formula was given in equation (\ref{Fj}). In the special example of partially reactive surfaces, we also showed how to incorporate non-Markovian models of absorption using the encounter-based method.

One natural extension of the current work would be to consider non-spherical trap geometries along the lines of Ref. \cite{Coombs15} for totally absorbing traps. This  would require taking into account the intrinsic capacitance and dipole vectors  that are determined by the shape of $\calU_j$. In particular, the analysis of the inner solutions would need to be modified using a different stretched coordinate system. Another non-trivial extension would be to apply the encounter-based method to the full model given by equations (\ref{master}). This would involve both the absorption process within each trap along the lines of Ref. \cite{Bressloff22b,Bressloff22c} as well as the semipermeable interface itself. As we have recently shown elsewhere \cite{Bressloff23a,Bressloff23b}, the latter can be reformulated using a probabilistic formulation of Brownian motion through a semipermeable interface known as snapping out Brownian motion \cite{Lejay16}. The latter sews together successive rounds of partially reflected Brownian motion that are restricted to either the interior or exterior of each interface. Each round is killed when an associated local time exceeds a random threshold, analogous to the mechanism considered in section 7.

Finally, it is useful to relate our work to another class of problem involving semipermeable interfaces, namely, coupled PDE-ODE systems. Let us return to the full model system (\ref{master}) and assume that $D_j$ is sufficiently large so that the density $q_j$ within $\calU_j$ is approximately spatially uniform. This means that we can take
\begin{equation}
q_j(\x,t)\approx q_j(t) \equiv \frac{1}{|\calU_j|} \int_{\calU_j}q_j(\x,t)d\x.
\end{equation}
For convenience, we have dropped the explicit dependence on $\x_0$.
Integrating both sides of equation (\ref{master}b) with respect to $\x\in \calU_j$ and using the divergence theorem gives
\begin{equation}
\frac{dq_j(t)}{dt}=\frac{D_j}{|\calU_j|}\int_{\partial \calU_j}\nabla q_j(\x,t)\cdot \n_j d\x-\gamma q_j(t),
\end{equation}
Using the semipermeable boundary conditions (\ref{master}c), this becomes
\begin{subequations} 
\label{mastercomp}
\begin{equation}
\frac{dq_j(t)}{dt}=\frac{\kappa_j}{|\calU_j|}\int_{\partial \calU_j}\left ([1-\alpha_j] p(\x,t) -\alpha_j q_j(t)\right )d\x-\gamma q_j(t),
\end{equation}
with $p(\x,t)$ the solution to
\begin{align}
	\frac{\partial p(\x,t)}{\partial t} &= D\nabla^2 p(\x,t), \ \x\in \Omega\backslash \calU_a,\\
 D\nabla p(\x,t)\cdot \n_j &= \kappa_j [(1-\alpha_j )p(\x,t)- \alpha_jq_j(t)],\quad \x  \in \partial \calU_j ,\\
 \nabla p\cdot \n&=0,\ \x \in \partial \Omega .
\end{align}
\end{subequations}
Equations (\ref{mastercomp}) are an example of a PDE-ODE system that couples bulk diffusion in $\Omega\backslash \calU_a$ with a set of $N$ spatially uniform compartments. The coupling is mediated by an effective semipermeable boundary condition. More general versions of this type of model are obtained by replacing the decay term in the ODE (\ref{mastercomp}a) with a nonlinear function $f(q_j,{\bf w}_j)$, where ${\bf w}_j$ represents a set of $m$ additional variables that evolve according to a system of ODEs that in turn couple to $q_j$.

One major application of the nonlinear version of equations (\ref{mastercomp}) is bacterial quorum sensing \cite{Muller06,Muller13,Gou16}. Quorum sensing involves the production and extracellular secretion of signaling molecules known as autoinducers. Each cell contains
receptors that specifically detect the signaling molecule (inducer), which then activates transcription of certain genes, including 
those for inducer synthesis. However, since there is a low likelihood of an individual bacterium detecting its own secreted inducer, the cell must encounter 
signaling molecules secreted by other cells in its environment in order for gene transcription to be activated. The extracellular concentration depends on the bacterial population density. Hence, as the population grows, the concentration of the inducer passes a threshold, causing more inducer to be 
synthesized. This generates a positive feedback loop that fully activates the receptor, and induces the up-regulation of other specific genes. All of the cells thus
initiate transcription at approximately the same time, resulting in some form of coordinated behavior such as bioluminescence, biofilm formation or infection of an animal host.
With reference to (\ref{mastercomp}), $p(\x,t)$ would represent the extracellular autoinducer concentration, whereas $q_j(t)$ would represent the concentration inside the $j$th cell. On the other hand, the additional variables ${\bf w}_j$ would represent the concentrations of downstream signaling proteins and genes that are activated by the autoinducers within the $j$th cell. At the population level, one is typically interested in the steady-state solution and its stability, both of which can be analyzed using matched asymptotics \cite{Muller06,Muller13,Gou16}. These studies typically take $\alpha_j=1/2$ (unbiased interfaces).

\end{document}